\documentstyle[aps,multicol,tighten,eqsecnum,psfig]{revtex}

\begin{document}
\draft

\title{Generating entangled superpositions of macroscopically 
       distinguishable states \\ within a parametric oscillator}

\author{Francesco De Martini,$^{1,2}$ Mauro Fortunato,$^{2,3,}$\cite{mau}
Paolo Tombesi,$^{2,3}$ and David Vitali$^{2,3}$}
\address {$^1$Dipartimento di Fisica, Universit\`a ``La Sapienza''
I--00185 Roma, Italy \\
$^{2}$Istituto Nazionale di Fisica della Materia, Italy \\
$^3$~Dipartimento di Matematica e Fisica,
Universit\`a di Camerino, via Madonna delle Carceri I--62032 
Camerino, Italy}
\date{\today}
\maketitle
\begin{abstract}
We suggest a variant of the recently proposed experiment for the generation
of a new kind of Schr\"odinger-cat states, using two coupled parametric
down-converter nonlinear crystals $\lbrack$F.~De~Martini, Phys. Rev. Lett. {\bf 
81}, 2842 (1998)$\rbrack$. We study the parametric oscillator case and find that
an entangled Schr\"odinger-cat type state of two cavities, whose mirrors are
placed along the output beams of the nonlinear crystals, can be realized 
under suitable conditions.
\end{abstract}


\begin{multicols}{2}

\narrowtext

\section{Introduction}
\label{intro}

Schr\"odinger-cat states~\cite{kn:sc,kn:cat} are most important in the domain of 
fundamental quantum mechanics, since the study of their progressive
decoherence~\cite{kn:zur,kn:prlha} would provide a better understanding of the
transition from the {\em quantum} to the {\em classical} 
world~\cite{kn:tra}. However, due to their extreme sensitivity to the
decoherence caused by the interaction with the environment, such 
linear superpositions of macroscopically distinguishable states are 
difficult to produce and to observe~\cite{kn:zur,kn:prlha}. In the 
last few years, a major effort in this field has led to the 
experimental production and detection of {\em mesoscopic} superpositions
of distinct states, both in the context of the single-mode microwave
cavities~\cite{kn:prlha} and of the dynamics of the
center of mass motion of a trapped 
ion~\cite{kn:win}. On the other hand, entanglement has been widely
recognized as one of the essential and most puzzling features of
quantum mechanics~\cite{kn:ent}, in that it allows the existence of
{\em quantum correlated} states of two noninteracting subsystems:
Entangled states play a crucial role in the so called Einstein,
Podolsky and Rosen (EPR) paradox~\cite{kn:epr}, and are essential in
the rapidly growing field of quantum information, as they allow the
feasibility of quantum state teleportation~\cite{kn:tel}, quantum
cryptography~\cite{kn:cryp}, and quantum computation~\cite{kn:qcomp}.

In two recent papers~\cite{kn:dem}, one of us has proposed an
original scheme for the generation of a new kind of {\em amplified}
Schr\"odinger-cat type states. It is based on the new concept of {\em 
quantum injection} into an optical parametric amplifier (OPA) operating in
{\em entangled} configuration.

\begin{figure}
\centerline{\psfig{figure=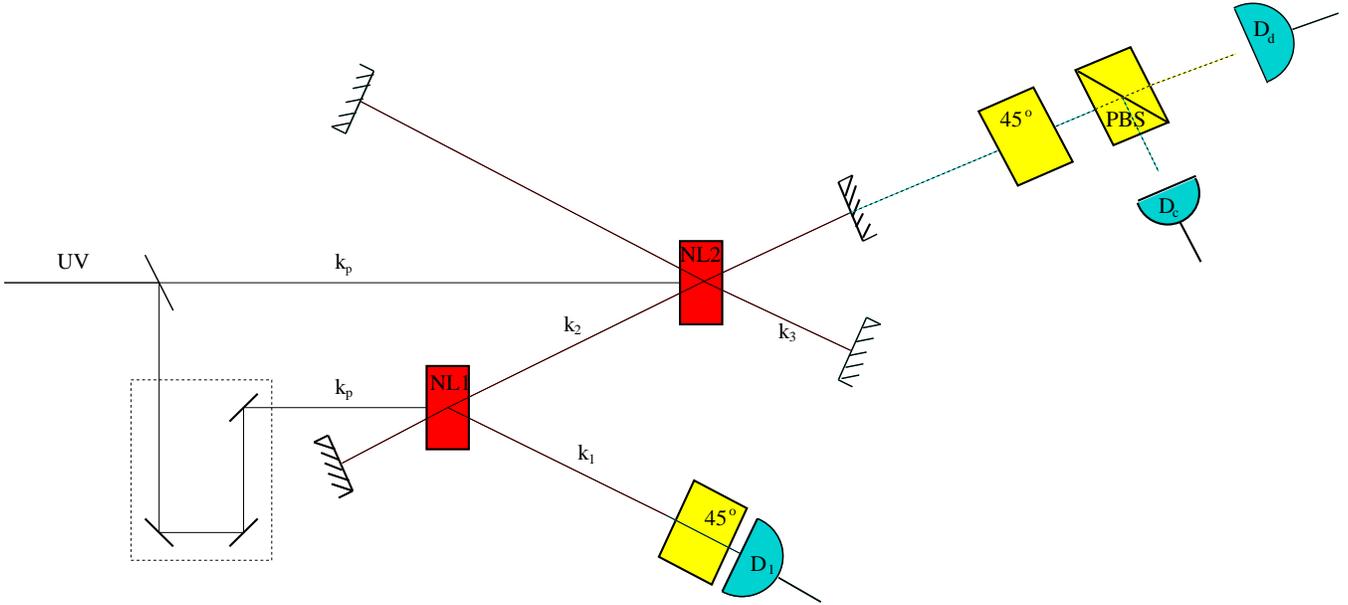,width=7in}}
\vspace{0.2cm}
\caption{\widetext
         Scheme of the experimental apparatus required for the 
         generation and detection of entangled superpositions of
         macroscopically distinguishable states: The idler beam 
         ($k_{2}$)
         of the first nonlinear crystal NL1 is used to inject a
         second nonlinear crystal NL2, while its signal beam ($k_{1}$) 
         triggers the photodetector D$_{1}$ after passing through a 
         polarizer. Modes $k_{2}$ and $k_{3}$ are placed within couples
         of mirrors. The detection apparatus---a rotator, a 
         polarizing beam splitter PBS and the two detectors D$_{c}$ 
         and D$_{d}$---probes the field along $k_{2}$, partially 
         leaking through one of the mirrors of the parametric 
         oscillator.}
\label{fg:scheme}
\end{figure}

As a relevant variant and a natural extension of the above scheme,
in the present work we analyze the case of the quantum injection in an optical 
parametric oscillator (OPO) in which two optical cavities are added to 
the OPA scheme considered in Ref.~\cite{kn:dem}: refer to Fig.~\ref{fg:scheme}.
Since the presence of the cavities 
leads to a large enhancement of the nonlinear (NL) parametric 
interaction, the number of the photon couples which are expected to 
be generated, in practical conditions, by the OPO scheme is far 
larger than in the amplifier condition: In addition, the generation of
parametrically coupled quasi-coherent fields represents in this 
context an appealing perspective.

The Schr\"odinger-cat state that has been put forward in 
Ref.~\cite{kn:dem} and is being analyzed, in a more detailed fashion, in the
present paper, is a superposition of
two macroscopic states which are distinguished by their polarization.
It can be considered as a sort of amplified version of the
polarization-entangled 
states which have been widely used in the last few years for the demonstration
of the violation of Bell's inequality~\cite{kn:kwiat,kn:bell}, of 
teleportation~\cite{kn:tel}, and for the generation of 
Greenberger-Horne-Zeilinger (GHZ) states~\cite{kn:ghz}.

The present paper is organized as follows:
In Sec.~\ref{spdc} we 
briefly describe the process of type-II parametric down conversion,
with an emphasis on the kind of entangled states usually produced in 
these experiments, and on the state we want to generate.
In Sec.~\ref{scheme} we outline the experimental apparatus needed for 
the realization of our scheme.
We devote Sec.~\ref{evol} to the presentation of the dynamical time 
evolution of the density matrix and of the Wigner function in our 
system,
and Sec.~\ref{stability} to the discussion of the stability 
conditions for our parametric oscillator.
In Sec.~\ref{choice} we set the initial conditions for the two coupled
nonlinear crystals and the two cavities,
whereas the way in which the 
cat state is produced is discussed in detail in Sec.~\ref{genera}.
Sec.~\ref{detect} is devoted to the presentation of the three methods we 
propose for detecting the Schr\"odinger-cat state: photodetection
(Sec.~\ref{photo}), measurement of the second-order quantum coherence
(Sec.~\ref{corr}), and Wigner function reconstruction  (Sec.~\ref{wigner}).
We finally summarize and discuss our results in Sec.~\ref{conclu}. The
appendix is devoted to the development of the small interaction time 
approximation.

\end{multicols}

\widetext

\begin{multicols}{2}

\section{Entanglement generating Parametric Down Conversion}
\label{spdc}

Let us first describe the kind of states commonly generated in the experiments
aimed at the violation of the Bell's inequalities.
In these experiments the NL crystal (typically beta-barium-borate: 
BBO) is cut for Type II phase matching where the two down-converted 
photons are emitted into two cones, one ``ordinary'' polarized
($o$), the other ``extraordinary'' polarized ($e$). When the angle between the
pump direction and the nonlinear crystal optical axis is sufficiently
large~\cite{kn:kwiat}, the two cones mutually intersect along two lines,
lying on opposite sides of the pump beam direction. These ones 
identify the output modes of the parametric down conversion:
$\vec{k}_{j}$ ($j=1$, $2$).
Therefore the field belonging to the modes $\vec{k}_{j}$ can be 
simultaneously $e$- and $o$-polarized. In typical conditions, the 
output state of the emitted photon couple may be expressed 
by~\cite{kn:tel,kn:dem,kn:shih}
\begin{equation}
|\psi \rangle =\frac{1}{\sqrt{2}}\left(|e_{1},o_{2}\rangle +e^{i\phi
}|o_{1}, e_{2}\rangle \right) \;.
\label{eq:pola}
\end{equation}
Since we have, for each couple, four degrees of freedom involved,
i.e. 2 states of orthogonal linear polarization $e$, $o$ for each
mode $\vec{k}_{j}$, we can rewrite state (\ref{eq:pola}) in the more
precise form 
\begin{equation}
|\psi \rangle =\frac{1}{\sqrt{2}}\left(|1\rangle _{1e} |1\rangle _{2o}
|0\rangle _{2e} |0\rangle _{1o}+e^{i\phi }|1\rangle _{1o} |1\rangle _{2e}
|0\rangle _{1e} |0\rangle _{2o} \right) \;,
\label{eq:pola2}
\end{equation}
which will be used in the following.

The ``Schr\"odinger-cat state'' we want to generate is a sort of 
amplification of this state, that is, it may be expressed in the form
\begin{eqnarray}
|\psi \rangle  & = & \frac{1}{\sqrt{2}}\left(|\psi ^{N}\rangle _{1e} 
|\psi ^{N}\rangle _{2o}
|0\rangle _{2e} |0\rangle _{1o} \right.
\nonumber \\
& & \left. +e^{i\phi}|\psi ^{N} \rangle _{1o} 
|\psi ^{N} \rangle _{2e} |0\rangle _{1e} |0\rangle _{2o} \right) \;,
\label{eq:catpro}
\end{eqnarray}
where $|\psi ^{N}\rangle$ is a state with a large number of photons in
some sense, and the states $|0\rangle$ are to be interpreted here as
squeezed vacuum states. This kind of state is different from the 
traditional Schr\"odinger cat states discussed in the quantum optics 
literature \cite{kn:zur,kn:prlha}, where one has a {\it single} mode of the
electromagnetic field in a superposition of two macroscopic states with
different phases of the 
field. The state (\ref{eq:catpro}) is a {\em nonlocal} superposition in which a 
macroscopic optical field is ``localized'' simultaneously either in the 
$e$- or in the $o$-polarized mode. In other words it is a state more 
similar to nonlocal field states such as
\begin{equation}
|\psi \rangle =\frac{1}{\sqrt{2}}\left(|\alpha \rangle _{1} 
|0\rangle _{2} + |0\rangle _{1} |\alpha \rangle _{2} \right) \;,
\label{eq:catloc}
\end{equation}
where the field can be simultaneously in one cavity or in another cavity 
and whose generation is discussed in \cite{kn:haro}.

We shall present here an experimental scheme for the generation of a state
which is actually a mixed state, but nonetheless, has the same structure of
the state of Eq.~(\ref{eq:catpro}), that is, can be represented by the
density operator
\begin{eqnarray}
\rho & = & \frac{1}{2}\left(\rho (N)_{1e,2o} \otimes 
\rho (0) _{1o,2e}+\rho (0)_{1e,2o} \otimes 
\rho (N) _{1o,2e} \right.
\nonumber \\
& & \left. + \rho ({\rm INT})_{1e,2o} \otimes 
\rho ({\rm INT'}) _{1o,2e}\right.
\nonumber \\
& & \left. +\rho ({\rm INT})_{1e,2o} ^{\dagger}\otimes 
\rho ({\rm INT'})^{\dagger} _{1o,2e}\right)\;,
\label{eq:catpro2}
\end{eqnarray}
where $\rho (N)$ is a two-mode mixed state with a large number of 
photons, $\rho (0)$ is a two-mode mixed state with a small number of 
photons and $\rho ({\rm INT})$ and $\rho ({\rm INT'})$ are the
interference terms.

\section{The experimental scheme}
\label{scheme}

We shall consider an experimental arrangement, Fig.~\ref{fg:scheme},
based on the one proposed in Ref.~\cite{kn:dem} and similar to that 
adopted in Refs.~\cite{kn:mandel} to show the realization of 
inducing coherence, without induced emission. Two down-conversion
NL crystals, are arranged in such a way that the two corresponding idlers 
beams are aligned along a common direction $\vec{k}_{2}$. 
Moreover, both idler beams and the signal beam
of one NL crystal (with wave-vector $\vec{k}_{3}$) are placed within 
couples of mirrors. This scheme can be thought of to realize the coupling
of two nondegenerate OPOs. The signal beam of the other crystal,
emitted along the direction $\vec{k}_{1}$ triggers the photodetector
${\rm D_{1}}$.

The directions $\vec{k}_{1}$, $\vec{k}_{2}$ and $\vec{k}_{3}$ are
selected to realize for both NL crystals the Type-II 
phase matching described before. These beams are then associated with six 
modes, with annihilation operator $a_{1o}$, $a_{1e}$, $a_{2o}$, $a_{2e}$,
$a_{3o}$ and $a_{3e}$. Note that the first two annihilation operators 
refer to traveling-waves, while the last four refer to cavity 
modes.

The dynamics of the system is determined by the nonlinear parametric 
interaction at each crystal and by the damping terms associated   
with losses and dissipation inside the cavities~\cite{kn:leg2}, as we
shall see in the next section.

\section{Time evolution for the density matrix and the Wigner function}
\label{evol}

The partial Hamiltonian operators describing the unitary dynamics 
inside the crystals are given by~\cite{kn:parosc}
\begin{mathletters}
\label{eq:hnl}
\begin{eqnarray}
\hat{H}_{\rm NL1} & = & i\hbar\chi_{1}(\hat{a}_{1e}^{\dagger}
\hat{a}_{2o}^{\dagger}-\hat{a}_{1e}\hat{a}_{2o})
\nonumber \\
& & +i\hbar\chi_{1}
(\hat{a}_{1o}^{\dagger}\hat{a}_{2e}^{\dagger}-\hat{a}_{1o}\hat{a}_{2e})
\label{eq:hnl1}\;, \\
\hat{H}_{\rm NL2} & = & i\hbar\chi_{2}(\hat{a}_{2o}^{\dagger}
\hat{a}_{3e}^{\dagger}-\hat{a}_{2o}\hat{a}_{3e})
\nonumber \\
& & +i\hbar\chi_{2}
(\hat{a}_{2e}^{\dagger}\hat{a}_{3o}^{\dagger}-\hat{a}_{2e}\hat{a}_{3o})
\label{eq:hnl2}\;,
\end{eqnarray}
\end{mathletters}
where $\chi_{1}=\epsilon_{1}\chi^{(2)}$, $\chi_{2}=\epsilon_{2}\chi^{(2)}$,
$\chi^{(2)}$ is the second-order nonlinear susceptibility of the 
crystals, and $\epsilon_{i}$ ($i=1,2$) is the pump intensity in 
crystals 1 and 2, respectively, which is assumed to be ``classical''.

Due to the explicit presence of dissipation in this problem, one has to
write the master equation for the reduced density matrix of the 
combined system which arises from the Hamiltonian terms 
(\ref{eq:hnl1}) and (\ref{eq:hnl2}) and from the damping terms
\begin{equation}
{\cal L}_{i}\rho=\kappa_{i}(2\hat{a}_{i}\rho\hat{a}_{i}^{\dagger}
                 -\hat{a}_{i}^{\dagger}\hat{a}_{i}\rho
                 -\rho\hat{a}_{i}^{\dagger}\hat{a}_{i})\;,
\label{eq:damping}
\end{equation}
for $i=2e, 2o, 3e, 3o$. Since the damping constants $\kappa_{i}$ are 
essentially connected to the transmittivity of the mirrors, it is quite 
natural to assume $\kappa_{2e}=\kappa_{2o}=\kappa_{2}$ and
$\kappa_{3e}=\kappa_{3o}=\kappa_{3}$.

Upon writing the full master equation for the total density matrix 
$\rho_{\rm T}$ of the (six-mode) system, it appears clear that the dynamics
of the six modes actually decouples into two independent dynamics for
two groups of three modes. In fact, one has
\begin{equation}
\dot{\rho}_{\rm T}= {\cal L}_{1e-2o-3e}\rho_{\rm T}
                   +{\cal L}_{1o-2e-3o}\rho_{\rm T}\;,
\label{eq:meq}
\end{equation}
where
\begin{eqnarray}
{\cal L}_{1e-2o-3e}\rho_{\rm T} & = &
-\frac{i}{\hbar}[\hat{H}_{1e-2o-3e},\rho_{\rm T}]
\label{eq:parmeq} \\
& + & \kappa_{2}(2\hat{a}_{2o}\rho_{\rm T}\hat{a}_{2o}^{\dagger}
                 -\hat{a}_{2o}^{\dagger}\hat{a}_{2o}\rho_{\rm T}
                 -\rho_{\rm T}\hat{a}_{2o}^{\dagger}\hat{a}_{2o})
\nonumber \\
& + & \kappa_{3}(2\hat{a}_{3e}\rho_{\rm T}\hat{a}_{3e}^{\dagger}
                 -\hat{a}_{3e}^{\dagger}\hat{a}_{3e}\rho_{\rm T}
                 -\rho_{\rm T}\hat{a}_{3e}^{\dagger}\hat{a}_{3e})\;,
\nonumber
\end{eqnarray}
and
\begin{eqnarray}
\hat{H}_{1e-2o-3e} & = & i\hbar\chi_{1}(\hat{a}_{1e}^{\dagger}
\hat{a}_{2o}^{\dagger}-\hat{a}_{1e}\hat{a}_{2o})
\nonumber \\
& & +i\hbar\chi_{2}
(\hat{a}_{2o}^{\dagger}\hat{a}_{3e}^{\dagger}-\hat{a}_{2o}\hat{a}_{3e})
\label{eq:hnl123}\;.
\end{eqnarray}
${\cal L}_{1o-2e-3o}$ is identical to ${\cal L}_{1e-2o-3e}$ up to the
substitution $e\to o$ and $o\to e$. As a consequence, the complete time
evolution will be of the form
\begin{equation}
\rho_{\rm T}(t)=e^{{\cal L}_{1e-2o-3e}t}e^{{\cal L}_{1o-2e-3o}t}
                \rho_{\rm T}(0)\;.
\label{eq:timev}
\end{equation}
From Eq.~(\ref{eq:timev}) it is clear that if the initial condition is
factorized, namely, if
\begin{equation}
\rho_{\rm T}(0)=\rho_{1e-2o-3e}(0)\otimes \rho_{1o-2e-3o}(0)\;,
\label{eq:factin}
\end{equation}
the state will remain factorized at all times, unless specifically 
designed conditional measurements~\cite{kn:cond} are performed on the system
(for example on the mode $\vec{k}_{1}$).

Due to the decoupling between the $1e-2o-3e$ and the $1o-2e-3o$
modes, we can simply restrict ourselves to the investigation of
the three-mode problem described by the master equation (\ref{eq:meq})
with (\ref{eq:parmeq}), and we shall drop the subscript $e$ and $o$
when not needed.

The Wigner function~\cite{kn:wig} 
$W(x_{1},y_{1},x_{2},y_{2},x_{3},y_{3})=W(\alpha_{1},\alpha_{2},\alpha_{3})$,
with $\alpha_{i}=x_{i}+iy_{i}$ $(i=1,2,3)$, resulting from this density matrix 
$\rho$ will then be a function of six real variables (or three complex
variables). Its time evolution, upon evaluating the commutator and the 
damping terms and after some lengthy algebra, is described by the
six-dimensional Fokker-Planck equation
\begin{equation}
\frac{\partial}{\partial t} W(\vec{z},t) = 
\gamma_{ij}\frac{\partial}{\partial z_{i}} 
\left(z_{j}W(\vec{z},t)\right) + D_{ij}\frac{\partial}{\partial 
z_{i}\partial z_{j}}W(\vec{z},t)\;,
\label{eq:6fp}
\end{equation}
where the vector $\vec{z}=(x_{1},y_{1},x_{2},y_{2},x_{3},y_{3})$, the 
matrix $D={\rm diag}(0,0,\kappa_{2}/4,\kappa_{2}/4,\kappa_{3}/4,\kappa_{3}/4)$, and
\begin{equation}
\gamma=\left(\matrix{0&0&-\chi_{1}&0&0&0 \cr
                     0&0&0&\chi_{1}&0&0 \cr
                     -\chi_{1}&0&\kappa_{2}&0&-\chi_{2}&0 \cr
                     0&\chi_{1}&0&\kappa_{2}&0&\chi_{2} \cr
                     0&0&-\chi_{2}&0&\kappa_{3}&0 \cr
                     0&0&0&\chi_{2}&0&\kappa_{3}}\right)\;.
\label{eq:gamma6}
\end{equation}

The solution to Eq.~(\ref{eq:6fp}) can be written~\cite{kn:gar} as the 
integral
\begin{equation}
W(\vec{z},t)=\int d^{4}z'\, W(\vec{z}\;',0) T(\vec{z},\vec{z}\;',t)\;,
\label{eq:solfp}
\end{equation}
where
\begin{mathletters}
\label{eq:where}
\begin{eqnarray}
T(\vec{z},\vec{z}\;',t) & = & \frac{1}{(2\pi)^{3}}
\frac{1}{\protect\sqrt{{\rm Det} \sigma (t)}}
\label{eq:tzzt}\\
 & \times & 
\exp\left[-\frac{1}{2}\langle\vec{z}-G(t)\vec{z}\;'|\sigma^{-1}(t)|\vec{z}
-G(t)\vec{z}\;'\rangle\right]\;,
\nonumber \\
G(t) & = & \exp\left(-\gamma t\right)\;,
\label{eq:gt}
\end{eqnarray}
and
\begin{equation}
\sigma(t)=2\int\limits_{0}^{t}d\tau\, G(\tau) D G^{t}(\tau)\;,
\label{eq:sigma}
\end{equation}
\end{mathletters}
$G^{t}$ being the transposed of the matrix $G$.

\section{Stability}
\label{stability}

The stability properties of the system are intimately connected to the
threshold of the overall OPO consisting of NL1 and NL2. Below threshold,
the system is stable and 
reaches a stationary state, since all eigenvalues of $\gamma$ have
positive real parts. On the other hand, above threshold the system is
unstable and its energy exponentially increases, because some 
eigenvalues of $\gamma$ have negative real part.

This result can be easily checked in the case in which the parametric
oscillator associated with NL2 is decoupled from NL1 ($\chi_{1}=0$):
in this case modes $\vec{k}_{2}$ and $\vec{k}_{3}$ decouple
from mode $\vec{k}_{1}$, and we end up with a four-dimensional
problem for the modes $\vec{k}_{2}$ and $\vec{k}_{3}$, described
by a Fokker-Planck equation of the same type as Eq.~(\ref{eq:6fp}),
but with
\begin{equation}
\gamma=\left(\matrix{\kappa_{2}&0&-\chi_{2}&0 \cr
                     0&\kappa_{2}&0&\chi_{2} \cr
                     -\chi_{2}&0&\kappa_{3}&0 \cr
                     0&\chi_{2}&0&\kappa_{3}}\right)\;,
\end{equation}
and $D={\rm diag}(\kappa_{2}/4,\kappa_{2}/4,\kappa_{3}/4,\kappa_{3}/4)$. In this case the 
(doubly degenerate) eigenvalues of $\gamma$ are
\begin{equation}
\lambda_{\pm}=\frac{\kappa_{2}+\kappa_{3}}{2}\pm\sqrt{\left(\frac{\kappa_{2}-\kappa_{3}}{2}
\right)^{2}+\chi_{2}^{2}}\;,
\label{eq:eigenv}
\end{equation}
and the stability condition becomes
\begin{equation}
\chi_{2}^{2}\leq \kappa_{2}\kappa_{3}\;,
\label{eq:stability}
\end{equation}
which coincides with the customary threshold for the parametric 
oscillator~\cite{kn:parosc}. However, if we turn on the first parametric
amplifier ($\chi_{1}\neq 0$) then the problem turns from 4-dimensional
to 6-dimensional, as we have seen: the eigenvalues of $\gamma$ change
and it is in principle possible to change the threshold, i.e.,
the stability condition.
As soon as $\chi_{1}\neq 0$, namely the first parametric amplifier is 
present, the system becomes unstable, independently on the values 
of $\chi_{2}$, $\kappa_{2}$, and $\kappa_{3}$. In fact, the eigenvalue equation 
for $\gamma$ is
\begin{equation}
(\lambda^{3}-(\kappa_{2}+\kappa_{3})\lambda^{2}
+(\kappa_{2}\kappa_{3}-\chi_{1}^{2}-\chi_{2}^{2})\lambda
+\kappa_{3}\chi_{1}^{2})^{2} = 0\;.
\label{eq:eigenunst}
\end{equation}
As a consequence, we have three doubly degenerate eigenvalues 
($\lambda_{1}$, $\lambda_{2}$, and $\lambda_{3}$). Since
$\lambda_{1}\lambda_{2}\lambda_{3}=-\kappa_{3}\chi_{1}^{2}$, at least one
of the $\lambda_{i}$ has a negative real part.

\section{Choice of the initial condition}
\label{choice}

We assume that at the beginning the first crystal is switched off (the
pump strength $\epsilon_{1}=0$). On the other hand, the second pump is
on ($\epsilon_{2}\neq 0$) and the second parametric oscillator is in 
its equilibrium state below threshold.
We therefore have a factorized initial state
\begin{equation}
\rho_{\rm T}(0) = \rho_{1e-2o-3e}(0)\otimes\rho_{1o-2e-3o}(0)\;,
\label{eq:infact}
\end{equation}
where
\begin{eqnarray}
\rho_{1e-2o-3e}(0) & = & \rho_{1o-2e-3o}(0)=\rho_{1-2-3}(0)
\nonumber \\
 & = & |0\rangle_{1} 
{}_{1}\langle 0| \otimes \rho_{2-3}(0)\;,
\label{eq:rho123}
\end{eqnarray}
and $\rho_{2-3}(0)$ is the equilibrium state of the oscillator below 
threshold. This can be easily determined upon considering the limits
\begin{equation}
\lim_{t \to \infty} G(t)=0\;, \,\,\,\,\,\,\, \lim_{t \to \infty} \sigma (t)
= \sigma(\infty)\;,
\label{eq:limit}
\end{equation}
and results in the following expression
\begin{eqnarray}
W(\vec{z}, t=\infty) & = & \int d^{4}z' W(\vec{z}\;',0) T(\vec{z}, 
\vec{z}\;', \infty)
\label{eq:weqbt} \\
 & = & \frac{1}{(2\pi)^{2}}\frac{1}{\protect\sqrt{\det 
 \sigma(\infty)}}
 \exp \left[ -\frac{1}{2}\vec{z}\sigma^{-1}(\infty) \vec{z}\right]\;.
\nonumber
\end{eqnarray}
The equilibrium state is thus a Gaussian state in which the modes $\vec{k}_{2}$
and $\vec{k}_{3}$ are correlated.

The initial state $\rho_{2-3}(0)$ is then given by the density matrix 
corresponding to the Wigner function
\begin{eqnarray}
W^{2-3}_{\rm bt}(0) & = & \left(\frac{2}{\pi}\right)^{2}
\left(1-\frac{\chi_{2}^{2}}{k^{2}}\right)
\nonumber \\
 & & \times \exp\left\{-2\left(x_{2}^{2}+y_{2}^{2}+x_{3}^{2}+y_{3}^{2}
 -2\frac{\chi_{2}}{k}
x_{2}x_{3}\right.\right.
\nonumber \\
 & & \left.\left. \,\,\,\,\,\,\,\,\,\,\,\,\,\,\,\,\,\,\,
 + 2\frac{\chi_{2}}{k} y_{2}y_{3}\right)\right\}\;,
\label{eq:w23bt}
\end{eqnarray}
where it is straightforward to realize that the modes 2 and 3 are 
correlated. Moreover, it is not a pure state, because
\begin{eqnarray}
{\rm Tr}(\rho^{2}) & = & \pi \int dx_{2} dy_{2} dx_{3} dy_{3}
\left[W_{\rm bt}^{2-3} (0)\right]^{2}
\nonumber \\
 & = & \left(1-\frac{\chi_{2}^{2}}{k^{2}}\right) < 1 \;,
\label{eq:trrhosq}
\end{eqnarray}
as expected.

The reduced density matrices of each mode are identical and coincide 
with the thermal state
\begin{equation}
W_{\rm bt}^{\rm red}(0) = \frac{2}{\pi}
\left(1-\frac{\chi_{2}^{2}}{k^{2}}\right)
\exp\left\{-2|\alpha|^{2}\left(1-\frac{\chi_{2}^{2}}{k^{2}}\right)\right\}\;,
\label{eq:redtherm}
\end{equation}
with an initial mean number of photons given by
\begin{equation}
\bar{N}=\frac{\chi_{2}^{2}}{2(k^{2}-\chi_{2}^{2})}\;,
\label{eq:meannumb}
\end{equation}
which means that when the oscillator is initially sufficiently close 
to threshold the initial mean number of photons in modes 2 and 3 
within the cavities can be large.

\section{Generation of the cat state}
\label{genera}

At time $t=0$ the first pump is turned on ($\epsilon_{1}\neq 0$): Also 
the first crystal begins to operate and the two groups of three modes
start their joint evolution, according to
\begin{eqnarray}
\rho_{\rm T}(t) & = & e^{{\cal L}_{1e-2o-3e}t}\rho_{1e-2o-3e}(0)
\nonumber \\
 & & \otimes e^{{\cal L}_{1o-2e-3o}t}\rho_{1o-2e-3o}(0)\;,
\label{eq:evolrhot}
\end{eqnarray}
where the two factorized evolutions are identical because both the
operator ${\cal L}$ and the initial condition are identical in the 
two cases. As a consequence, we end up with two identical 
six-dimensional problems.

The solution of Eq.~(\ref{eq:evolrhot}) can be found as in 
Sec.~\ref{evol} by using the Wigner functions
\begin{equation}
W_{123}(\vec{z}, t) = \int d^{6}z' W_{123}(\vec{z}\;',0)
T(\vec{z}, \vec{z}\;', t)\;,
\label{eq:wig123}
\end{equation}
where the initial Wigner function $W_{123}(\vec{z},0)$ corresponding
to the initial density matrix [Eq.~(\ref{eq:rho123})] is given by
\begin{mathletters}
\label{eq:wwww}
\begin{eqnarray}
W_{123}(\vec{z}, 0) & = & \left(\frac{2}{\pi}\right)^{3}
\left(1-\frac{\chi_{2}^{2}}{k^{2}}\right)
e^{-2(x_{1}^{2}+y_{1}^{2})}
\nonumber \\
 & & \times \exp\Big\{-2\Big[x_{2}^{2}+y_{2}^{2}+x_{3}^{2}+y_{3}^{2}
 \nonumber \\
 & & \,\,\,\,\,\,\,\,\,\,\,\,\,\,\,\,\,\,\,\,\, -2\left.
 \left.\frac{\chi_{2}}{k}\left(x_{3}x_{2}+y_{3}y_{2}
 \right)\right]\right\}
\label{eq:w123in} \\
 & = & \left(\frac{2}{\pi}\right)^{3}\sqrt{\det C}
\exp\left\{-2\langle\vec{z}|C|\vec{z}\rangle\right\}\;,
\label{eq:wcomp}
\end{eqnarray}
\end{mathletters}
with
\begin{equation}
C= \left(\matrix{
1 & 0 & 0 & 0 & 0 & 0 \cr
0 & 1 & 0 & 0 & 0 & 0 \cr
0 & 0 & 1 & 0 & -\frac{\chi_{2}}{k} & 0 \cr
0 & 0 & 0 & 1 & 0 & \frac{\chi_{2}}{k} \cr
0 & 0 & -\frac{\chi_{2}}{k} & 0 & 1 & 0 \cr
0 & 0 & 0 & \frac{\chi_{2}}{k} & 0 & 1 \cr
}\right)\;,
\label{eq:cmatrix}
\end{equation}
and
\begin{equation}
\det C = \left(1-\frac{\chi_{2}^{2}}{k^{2}}\right)^{2}\;.
\label{eq:detc}
\end{equation}

From Eq.~(\ref{eq:wig123}) one can immediately recognize that since 
the initial state $W_{123}(\vec{z}, 0)$ is Gaussian and the propagator
$T(\vec{z}, \vec{z}\;', t)$ is also Gaussian, the Wigner function
$W_{123}(\vec{z}, t)$ of the evolved state must remain Gaussian at all
times.  

Upon integrating over $d^{4}z'$, Eq.~(\ref{eq:wig123}) can be 
rewritten as
\begin{equation}
W(\vec{z},t)=\frac{\protect\sqrt{\det B(t)}}{\pi^{3}}
\exp \left\{-\langle\vec{z}|B(t)|\vec{z}\rangle\right\}\;,
\label{eq:sol123}
\end{equation}
where
\begin{equation}
B(t)=\left[2\sigma(t)+\frac{G(t)C^{-1}G^{t}(t)}{2}\right]^{-1}\;,
\label{eq:bt}
\end{equation}
and $G(t)$ and $\sigma(t)$ are the six-dimensional matrices defined 
in Eqs.~(\ref{eq:gamma6}), (\ref{eq:gt}), and (\ref{eq:sigma}).

This Gaussian evolution holds for a short time only. As a matter of 
fact, one should distinguish between the mode along direction 1 and
those along directions 2 and 3: $a_{2}^{\dagger}$ and $a_{3}^{\dagger}$
denote creation of a photon in the {\em stationary-wave} modes within the 
cavities, whereas $a_{1}^{\dagger}$ denotes the creation of a photon
in the {\em traveling-wave} mode along direction $\vec{k}_{1}$.
Therefore the interaction $H_{\rm NL1}=i\hbar\chi_{1}(a_{1}^{\dagger}
a_{2}^{\dagger}-a_{1}a_{2})$ exists only for the time period during 
which this traveling wave mode 1 moves {\em within} the nonlinear 
crystal. In order to prepare the desired state for the modes 2 and 3,
simultaneously taking full advantage of the degree of freedom 
represented by the traveling-wave mode 1, we perform a {\em 
conditional}~\cite{kn:cond} measurement on direction 1, thereby
conditioning the state of the four modes along directions 2 and 3
upon the detection of a photon along direction 1 polarized at $\pi/4$
with respect to the two output polarizations $e$ and $o$, which are
orthogonal to each other. In this way we also post-select (along
direction 2) the input state of the second crystal.
The projection operator associated to such a conditional measurement 
is therefore given by
\begin{eqnarray}
\hat{P}_{\frac{\pi}{4}} & = & \frac{1}{2}\left\{|1\rangle_{1o}|0\rangle_{1e}
+ |0\rangle_{1o}|1\rangle_{1e}\right\}
\nonumber \\
 & & \times \left\{{}_{1o}\langle 1|{}_{1e}\langle 0|
+ {}_{1o}\langle 0|{}_{1e}\langle 1|\right\}\;.
\label{eq:ppi4}
\end{eqnarray}
As a consequence of this measurement (whose success probability 
amounts to 0.5) the state along direction 1 and directions 2 and 3
factorizes: The state along direction 1 is given by
\begin{equation}
|\psi\rangle_{1}=\frac{1}{\protect\sqrt{2}}\left\{|1\rangle_{1o}|0\rangle_{1e}
+ |0\rangle_{1o}|1\rangle_{1e}\right\}\;,
\label{eq:polph}
\end{equation}
which represents a photon polarized at $\pi/4$,  whilst the 
conditional state for directions 2 and 3 is represented by the 
density matrix
\begin{eqnarray}
\rho^{c}_{2o-3e-2e-3o}(t) & \propto &
\left\{{}_{1o}\langle 1|{}_{1e}\langle 0|
+ {}_{1o}\langle 0|{}_{1e}\langle 1|\right\}
\rho_{1o-2e-3o}(t)
\nonumber \\
 & & \otimes \rho_{1e-2o-3e}(t)
\left\{|1\rangle_{1o}|0\rangle_{1e}\right.
\nonumber \\
 & & \,\,\,\,\,\,\,\,\,\,\,\,\,\,\,\,\,\,\,\,\,\,\,\,\,\,\,\,\,\,\,\,
+ \left.|0\rangle_{1o}|1\rangle_{1e}\right\}\;,
\label{eq:rho2323}
\end{eqnarray}
which can be rewritten as
\begin{eqnarray}
\rho^{c}_{2o-3e-2e-3o}(t) & \propto & 
\left[\rho^{(1)}_{2e-3o}\rho^{(0)}_{2o-3e}
+\rho^{(0)}_{2e-3o}\rho^{(1)}_{2o-3e}\right.
\nonumber \\
 + & & \left.\rho^{({\rm int})}_{2e-3o}\rho^{({\rm int})\,\dagger}_{2o-3e}
+\rho^{({\rm int})\,\dagger}_{2e-3o}\rho^{({\rm 
int})}_{2o-3e}\right]\;,
\label{eq:rhocon}
\end{eqnarray}
where
\begin{mathletters}
\label{eq:rhoconiii}
\begin{eqnarray}
\rho^{(1)}_{2-3} & = & {}_{1}\langle 1|\rho_{1-2-3}(t)|1\rangle_{1}\;,
\label{eq:rhocon231} \\
\rho^{(0)}_{2-3} & = & {}_{1}\langle 0|\rho_{1-2-3}(t)|0\rangle_{1}\;,
\label{eq:rhocon230} \\
\rho^{({\rm int})}_{2-3} & = & {}_{1}\langle 1|\rho_{1-2-3}(t)|0\rangle_{1}\;.
\label{eq:rhocon23i}
\end{eqnarray}
\end{mathletters}
The state of Eq.~(\ref{eq:rhocon}) is of the same form of the desired 
state, Eq.~(\ref{eq:catpro2}), and is a linear superposition of distinguishable
states, as long as $\rho^{(1)}_{2-3}$ is well distinguishable from
$\rho^{(0)}_{2-3}$.

It should also be emphasized at this stage that 
the density matrix~(\ref{eq:rhocon}) directly corresponds to the 
Wigner function, Eq.~(2) of Ref.~\cite{kn:dem}, obtained in the OPA 
case. The similarity between the OPO and the OPA configurations is better 
brought about in the limit of small interaction times (see the 
appendix, where it is also shown that---in this limit---many
of our results are very similar to those obtained in the OPA 
case~\cite{kn:dem}). Roughly speaking, one should recover the OPA 
results from the OPO ones in the limit $\kappa \to \infty$, since this 
condition means absence of cavity mirrors. However, this 
correspondence does not hold exactly because the initial state in the 
OPO case (the state present in the cavity at $t=0$, when the first 
nonlinear crystal is switched on) is slightly different. This fact 
explains the differences between the OPA and the OPO, which 
result in a far larger effective number of photons in the latter case.

\section{Detection of the cat state}
\label{detect}

How can we probe the quantum state produced in this 
parametric-oscillator entangled configuration, and prove that it 
actually represents a Schr\"odinger-cat state? In order to do
this, one has to 
independently show that i) the state is indeed made out of two
{\em macroscopically distinct components}, that ii) these two components
exhibit {\em quantum interference}, so that the state can be considered as a
true {\em linear superposition} rather than a {\em statistical 
mixture}, and that iii) the ``separation'' between the two components 
{\em scales with a macroscopic or mesoscopic parameter}, usually the number 
of photons.
To achieve this goal, we propose three different and independent 
methods---which can be used either alternatively or 
simultaneously---as we shall explain in detail in the next three 
subsections. 

\subsection{Photodetection}
\label{photo}

Let us employ photon number measurements for the modes along 
direction 2, thereby collecting the photon-number distributions 
$P(n_{2o})$ and $P(n_{2e})$. We therefore consider the reduced
density matrix obtained by performing the trace on the state of 
Eq.~(\ref{eq:rhocon}), that is,
\begin{eqnarray}
\rho_{2e} & = & {\rm Tr}_{2o-3e-3o}\left\{\rho_{2e-3o-2o-3e}(t)\right\}
\nonumber \\
 & = & \frac{1}{2}\left\{{\rm Tr}_{3o}\left[\rho_{2e-3o}^{(1)}\right]
{\rm Tr}_{23}\left[\rho_{2-3}^{(0)}\right]\right.
\nonumber \\
 & & + \left.{\rm Tr}_{3o}\left[\rho_{2e-3o}^{(0)}\right]
{\rm Tr}_{23}\left[\rho_{2-3}^{(1)}\right]\right\} 
\left[P\left(\frac{\pi}{4}\right)\right]^{-1}\;,
\label{eq:rhored2e}
\end{eqnarray}
where $P(\pi/4)$ is the probability of finding one photon with 
polarization at $\pi/4$, that is,
\begin{eqnarray}
P\left(\frac{\pi}{4}\right) & = & {\rm Tr}_{1o-1e-2e-3o-2o-3e}
\left[\hat{P}_{\frac{\pi}{4}}\rho_{1o-2e-3o}(t)\right.
\nonumber \\
 & & \otimes \left.\rho_{1e-2o-3e}(t)\right]
\nonumber \\
 & = & {\rm Tr}_{2-3}\left[\rho_{2-3}^{(1)}\right]
 {\rm Tr}_{2-3}\left[\rho_{2-3}^{(0)}\right]\;,
\label{eq:probpi4}
\end{eqnarray}
represents the probability of the conditional measurement generating 
the desired cat state. The interference terms in Eq.~(\ref{eq:rhocon})
obviously give no contribution to Eq.~(\ref{eq:rhored2e}), since
\begin{equation}
{\rm Tr}_{2-3}\left[\rho_{2-3}^{({\rm int})}\right]=
{\rm Tr}_{2-3}\left[{}_{1}\langle 1 | \rho_{1-2-3}(t) | 0\rangle_{1}\right]
=0\;.
\label{eq:trint}
\end{equation}
Combining Eqs.~(\ref{eq:rhored2e}) and (\ref{eq:probpi4}), one obtains 
for the reduced state
\begin{equation}
\rho_{2e}=\frac{1}{2}\left[
\frac{{\rm Tr}_{3o}\rho_{2e-3o}^{(1)}}{{\rm Tr}_{2-3}\rho_{2-3}^{(1)}}
+\frac{{\rm Tr}_{3o}\rho_{2e-3o}^{(0)}}{{\rm Tr}_{2-3}\rho_{2-3}^{(0)}}
\right]\;,
\label{eq:2erhored}
\end{equation}
with an identical form for the reduced state $\rho_{2o}$. The reduced
density matrices $\rho_{3e}$ and $\rho_{3o}$ can be determined in a 
similar way.

\begin{figure}
\centerline{\psfig{figure=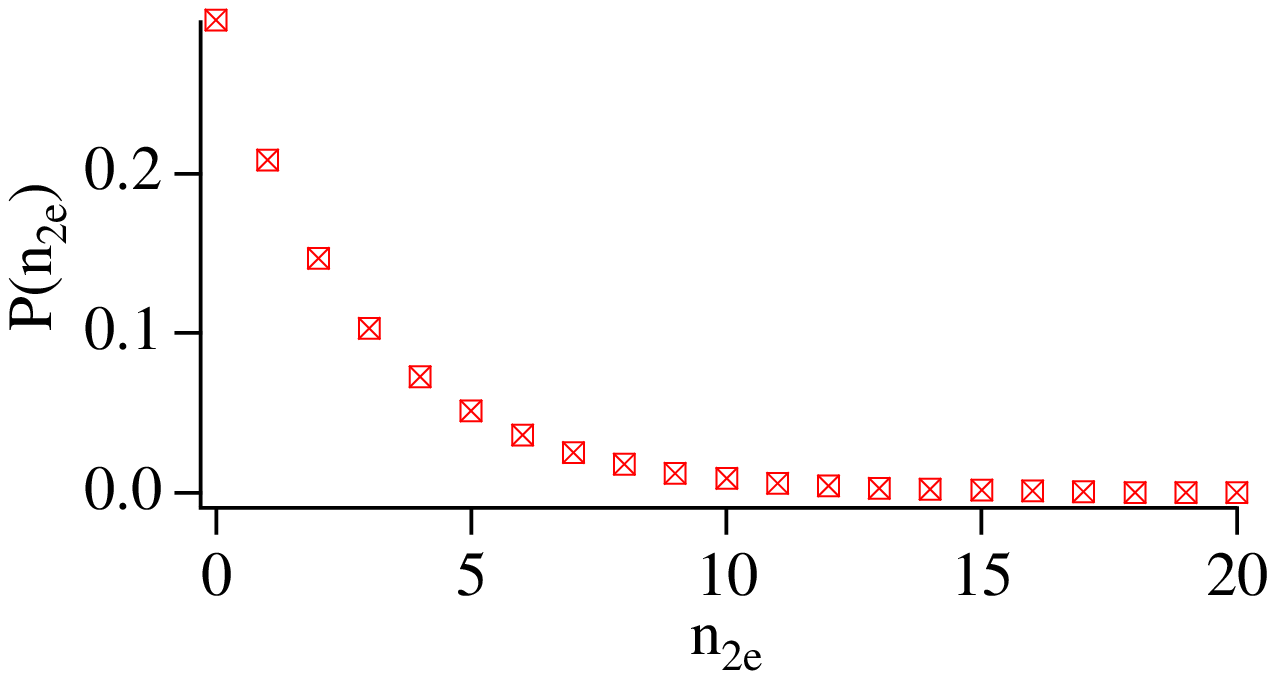,width=3.3in}}
\centerline{\psfig{figure=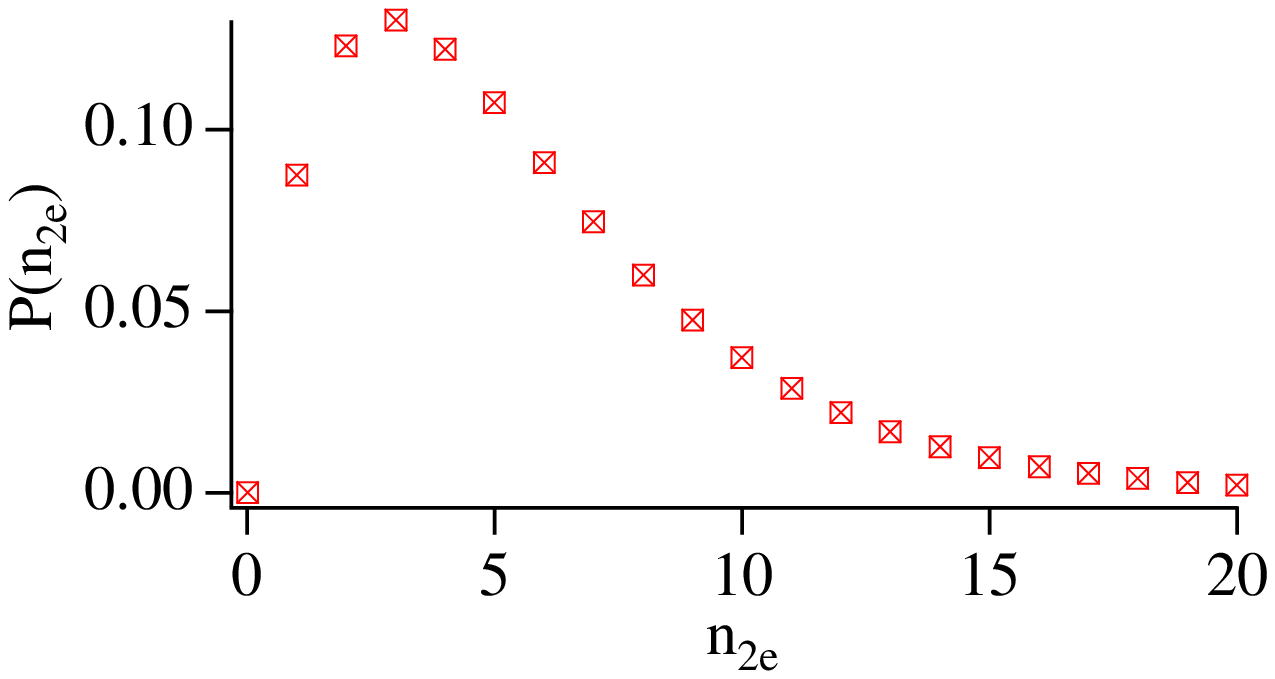,width=3.3in}}
\centerline{\psfig{figure=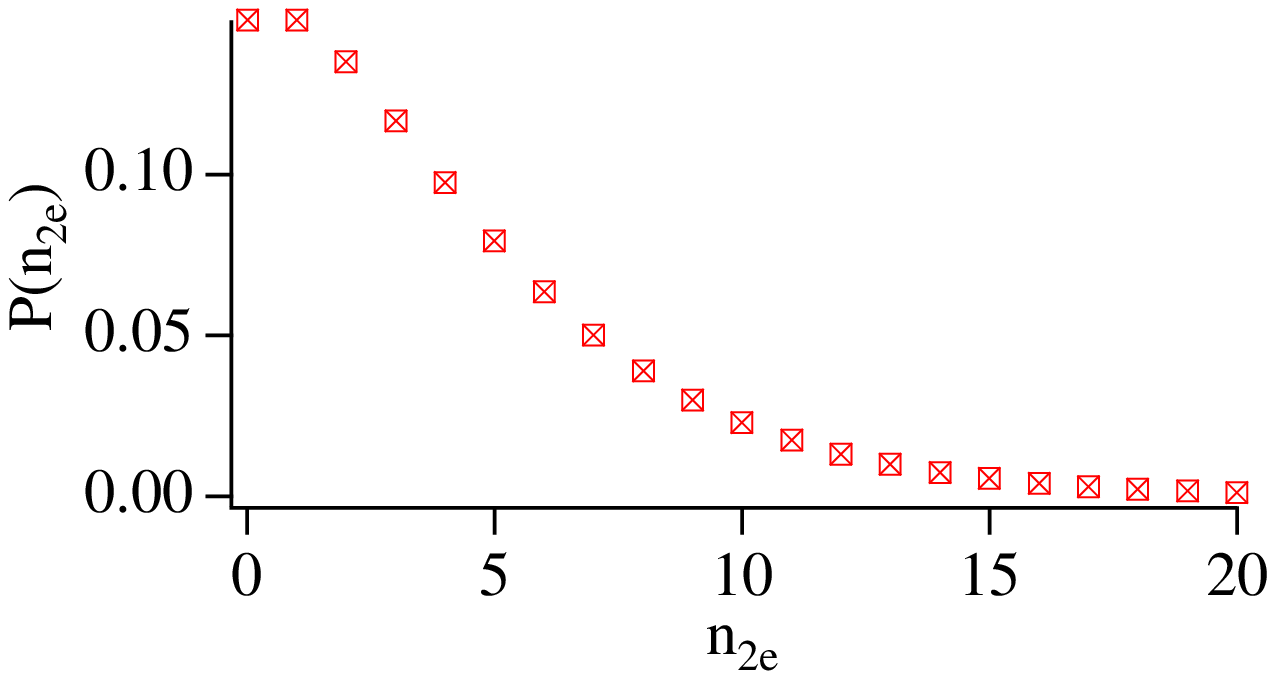,width=3.3in}}
\vspace{0.1cm}
\caption{\narrowtext
         Photon-number probability distributions for a photodetection
         experiment on mode $2e$. $P_{\rm H}(n_{2e})$, $P_{\rm 
         V}(n_{2e})$, and $P(n_{2e})$ [see Eq.~(\ref{eq:pn2e})] are
         plotted, respectively, in (a), (b), and (c).}
\label{fg:photod}
\end{figure}

From Eq.~(\ref{eq:2erhored}) it is immediate to recognize that the 
reduced state of the mode $2e$ is given by the sum of two density 
matrices, conditioned upon the detection of {\em one} photon
and of {\em zero} photons in the mode $1o$  (or, more precisely,
{\em one} photon in the mode $1e$), respectively. Therefore the two 
terms of the reduced density matrix can be experimentally obtained
by rotating the polarizer in front of the detector D$_{1}$ located along the
direction $\vec{k}_{1}$: When the polarizer is vertical (mode $1e$),
we have zero photons in the mode $1o$, and only the second term of 
the sum in the right-hand side (rhs) of Eq.~(\ref{eq:2erhored}) is realized.
On the contrary, if the polarizer is set horizontally (mode $1o$), one 
detects one photon in the mode $1o$, projecting the resulting
density matrix for the mode $2e$ onto the second term in the 
sum~(\ref{eq:2erhored}). However, both terms are present when the 
polarizer is set at 45$^{\circ}$. An experimentalist could then take 
advantage of this property to test the presence of the two component 
states: the distinction between the two states in the superposition
can be made via photon number measurements, yielding the probability 
distribution $P(n_{2e})$. In fact, one has
\begin{equation}
P(n_{2e})=\frac{1}{2}\left(P_{\rm H}(n_{2e})+P_{\rm 
V}(n_{2e})\right)\;,
\label{eq:pn2e}
\end{equation}
where $P_{\rm H}(n_{2e})$ [$P_{\rm V}(n_{2e})$] is the probability
distribution obtained when the polarized is set horizontally 
(vertically). The results an experimentalist would obtain with a
simple photodetection in these two situations are shown in 
Fig.~\ref{fg:photod}(a) and (b), together with the probability
distribution~(\ref{eq:pn2e}) one would obtain when the polarizer is
set at $45^{\circ}$ [Fig.~\ref{fg:photod}(c)].

In this way we have verified the existence of two distinct components 
in the state~(\ref{eq:2erhored}). But how can we be sure that these 
two components form a quantum superposition and not just a classical 
mixture? To answer this question, one has to perform  a measurement 
able to distinguish the ``cat state''
\begin{eqnarray}
\rho^{\rm cat}_{2o-3e-2e-3o}(t) & = & 
\left[2{\rm Tr}_{2-3}\rho_{2-3}^{(1)}{\rm 
Tr}_{2-3}\rho_{2-3}^{(0)}\right]^{-1}
\nonumber \\
 & \otimes & \left[\rho^{(1)}_{2e-3o}\rho^{(0)}_{2o-3e}
+ \rho^{(0)}_{2e-3o}\rho^{(1)}_{2o-3e}
\right.
\nonumber \\
 & + & \left.\rho^{({\rm int})}_{2e-3o}\rho^{({\rm int})\,\dagger}_{2o-3e}
     + \rho^{({\rm int})\,\dagger}_{2e-3o}\rho^{({\rm 
int})}_{2o-3e}\right]\;,
\label{eq:rhoconcomp}
\end{eqnarray}
from the corresponding statistical mixture
\begin{eqnarray}
\rho^{\rm mix}_{2o-3e-2e-3o}(t) & = & 
\left[2{\rm Tr}_{2-3}\rho_{2-3}^{(1)}{\rm 
Tr}_{2-3}\rho_{2-3}^{(0)}\right]^{-1}
\nonumber \\
 & \times & \left[\rho^{(1)}_{2e-3o}\rho^{(0)}_{2o-3e}
+ \rho^{(0)}_{2e-3o}\rho^{(1)}_{2o-3e}
\right]\;,
\label{eq:rhostat}
\end{eqnarray}
which does not exhibit any interference.

In order to reach this goal, we perform an interference experiment, 
involving the modes along direction $\vec{k}_{2}$ only, using a
detection system similar to the one proposed in Ref.~\cite{kn:dem},
as schematically described in Fig.~\ref{fg:scheme}. The measured 
quantity is given by the photocounts at the detector D$_{\rm c}$, as 
a function of the variable phase $\phi$. The annihilation operator $c$ 
corresponding to the mode traveling to the detector D$_{\rm c}$
can be written in terms of the annihilation operators of the modes
$2e$ and $2o$ as
\begin{equation}
c=\frac{1}{\protect\sqrt{2}}\left(a_{2o}+e^{i\phi}a_{2e}
\right)\;,
\label{eq:c}
\end{equation}
so that the operator number of photons for the mode $c$ will be given 
by
\begin{equation}
c^{\dagger}c=\frac{1}{2}\left(
a_{2o}^{\dagger}a_{2o}
+a_{2e}^{\dagger}a_{2e}
+e^{i\phi}a_{2o}^{\dagger}a_{2e}
+e^{-i\phi}a_{2e}^{\dagger}a_{2o}\right)\;.
\label{eq:nc}
\end{equation}
In order to be able to distinguish between the superposition state and 
the mixture, the expectation value
\begin{equation}
\langle c^{\dagger}c\rangle_{\rm cat} =
{\rm Tr}\left[c^{\dagger}c \rho_{2o-3e-2e-3o}^{\rm cat}(t)\right]
\label{eq:cdccond}
\end{equation}
has to be different from
\begin{equation}
\langle c^{\dagger}c\rangle_{\rm mix} =
{\rm Tr}\left[c^{\dagger}c \rho_{2o-3e-2e-3o}^{\rm mix}(t)\right]\;.
\label{eq:cdcmix}
\end{equation}
It is then clear that this interference experiment can answer our 
question whenever the contributions of the off-diagonal terms
${\rm Tr}[c^{\dagger}c \rho_{2e-3o}^{\rm (int)}
\rho_{2o-3e}^{\rm (int)\, \dagger}]$
and its complex conjugate
${\rm Tr}[c^{\dagger}c \rho_{2e-3o}^{\rm (int)\, \dagger}
\rho_{2o-3e}^{\rm (int)}]$ are nonzero.

Let us start by evaluating the contribution of the diagonal terms, 
namely, Eq.~(\ref{eq:cdcmix}). After explicit integration of the 
corresponding Wigner function, it is easy to prove that the 
phase-dependent terms [the third and the fourth term in 
Eq.~(\ref{eq:nc})] vanish when one computes the expectation value,
Eq.~(\ref{eq:cdcmix}). Therefore the diagonal terms yield a phase
($\phi$)-independent contribution given by
\begin{mathletters}
\label{eq:cdc}
\begin{eqnarray}
\langle c^{\dagger}c\rangle_{\rm mix} & = &
\frac{1}{2}\left(\langle a_{2o}^{\dagger}a_{2o}\rangle_{\rm mix}
+ \langle a_{2e}^{\dagger}a_{2e}\rangle_{\rm mix}\right)
\nonumber \\
 & = & \frac{1}{4}\left[\langle n_{2o}\rangle^{(1)}
+ \langle n_{2o}\rangle^{(0)}\right]
\nonumber \\
 & & + \frac{1}{4}\left[\langle n_{2e}\rangle^{(1)}
+ \langle n_{2e}\rangle^{(0)}\right]
\label{eq:cdcexp} \\
 & = & \frac{1}{2}\left[\langle n_{2}\rangle^{(1)}
+ \langle n_{2}\rangle^{(0)}\right]\;,
\label{eq:ncdiagexp}
\end{eqnarray}
\end{mathletters}
where
\begin{equation}
\langle n_{2}\rangle^{(i)} = 
\frac{{\rm Tr}_{2-3}\left[\rho_{2-3}^{(i)} a_{2}^{\dagger} 
a_{2}\right]}{{\rm Tr}_{2-3}\left[\rho_{2-3}^{(i)}\right]}\;,
\end{equation}
($i=0$, 1) is the mean photon number in one of the two diagonal states
in Eqs.~(\ref{eq:rhoconcomp}) and (\ref{eq:rhostat}). In the small 
interaction-time limit, which is very well justified in the present case
(see appendix), $1\gg kt$, $\chi_{1}t$, $\chi_{2}t$,
we have
\begin{mathletters}
\label{eq:23rho}
\begin{eqnarray}
\rho_{2-3}^{(0)} & \simeq & \rho_{2-3}(0)\;,
\label{eq:23rho0} \\
\rho_{2-3}^{(1)} & \propto & a_{2}^{\dagger}\rho_{2-3}(0) a_{2}\;.
\label{eq:23rho1}
\end{eqnarray}
\end{mathletters}
As a consequence, the two expectation values in the rhs of 
Eq.~(\ref{eq:ncdiagexp}) can be explicitly evaluated and are given by
\begin{mathletters}
\label{eq:n2}
\begin{eqnarray}
\langle n_{2} \rangle^{(0)} & = & \bar{N} =
\left[2\left(\protect\frac{k^{2}}{\chi_{2}^{2}} - 1 \right)\right]^{-1}\;,
\label{eq:n20} \\
\langle n_{2} \rangle ^{(1)} & = & 2\bar{N}+1\;,
\label{eq:n21}
\end{eqnarray}
\end{mathletters}
where $\bar{N}$ is the initial mean photon number in the cavity.
In conclusion, the diagonal contribution to 
the expectation value in Eq.~(\ref{eq:cdccond}) amounts to
\begin{equation}
\langle c^{\dagger} c \rangle_{\rm mix} \simeq \frac{1 + 3\bar{N}}{2}\;,
\label{eq:mixcdc}
\end{equation}
which is indeed $\phi$-independent as expected.

We turn now our attention to the off-diagonal terms in 
Eq.~(\ref{eq:rhoconcomp}), which are absent in Eq.~(\ref{eq:rhostat}).
First we note that the expectation values of the number operators 
relative to the two polarizations in mode 2 computed on the off-diagonal
terms vanish, i.e.,
\begin{equation}
\langle a_{2o}^{\dagger}a_{2o}\rangle_{\rm o-d}=
\langle a_{2e}^{\dagger}a_{2o}\rangle_{\rm o-d}=0\;.
\label{eq:numopod}
\end{equation}
On the other hand, the third and the fourth term in the rhs of 
Eq.~(\ref{eq:nc}) give to the expectation value on the off-diagonal 
terms the contributions
\begin{mathletters}
\label{eq:phio}
\begin{eqnarray}
e^{i\phi}\langle a_{2o}^{\dagger} a_{2e}\rangle_{\rm o-d} & = &
\left\{2{\rm Tr}_{2-3}\left[\rho_{2-3}^{(1)}\right]
{\rm Tr}_{2-3}\left[\rho_{2-3}^{(0)}\right]\right\}^{-1}
\nonumber \\
 & & \times \left(\langle a_{2e}\rangle^{\rm (int)}
 \langle a_{2o}^{\dagger}\rangle^{{\rm (int)}\, \dagger}\right.
\nonumber \\
 & &  \,\,\,\,\,\,\,\,\, + \left.\langle a_{2e}\rangle^{{\rm (int)}\, \dagger}
\langle a_{2o}^{\dagger}\rangle^{\rm (int)}\right)\;,
\label{eq:phiod} \\
e^{-i\phi} \langle a_{2e}^{\dagger} a_{2o} \rangle_{\rm o-d} & = &
\left(e^{i\phi}\langle a_{2o}^{\dagger} a_{2e}\rangle_{\rm o-d}\right)^{*}\;,
\label{eq:phiodst}
\end{eqnarray}
\end{mathletters}
where
\begin{equation}
\langle a_{2}\rangle^{\rm (int)}={\rm Tr}_{2-3}\left[\rho_{2-3}^{\rm (int)} 
a_{2}\right]\;.
\label{eq:a2int}
\end{equation}
These contributions are generally different from zero, and this 
observation is sufficient to reach the conclusion that the proposed interference 
experiment is able to distinguish the cat state from the corresponding
mixture.

We are able to evaluate these off-diagonal terms in the small 
interaction-time limit developed in the appendix: At the lowest order in
$\chi_{1}t$, $\chi_{2}t$, and $kt$, we have
\begin{equation}
\rho_{2-3}^{\rm (int)}\simeq a_{2}^{\dagger} \rho_{2-3}(0)\;,
\label{eq:intrho23}
\end{equation}
and therefore, using Eq.~(\ref{eq:a2int}),
\begin{mathletters}
\label{eq:inta2}
\begin{eqnarray}
\langle a_{2e}\rangle^{\rm (int)} & = & 
\chi_{1}t\left(\bar{N}+1\right)\;,
\label{eq:inta2e} \\
\langle a_{2e}\rangle^{{\rm (int)}\, \dagger} & = &
\chi_{1}t\langle a_{2e}^{2}\rangle = 0 \;,
\label{eq:inta2e+} \\
\langle a_{2o}^{\dagger}\rangle^{\rm (int)} & = &
\chi_{1}t \langle a_{2o}^{\dagger\, 2}\rangle = 0\;,
\label{eq:inta2o} \\
\langle a_{2o}^{\dagger}\rangle^{{\rm (int)}\, \dagger} & = &
\chi_{1}t\left(\bar{N}+1\right)\;.
\label{eq:inta2o+}
\end{eqnarray}
\end{mathletters}
On the other hand,
\begin{mathletters}
\label{eq:r23}
\begin{eqnarray}
\rho_{2-3}^{(1)} & \simeq & 
\chi_{1}^{2}t^{2}a_{2}^{\dagger}\rho_{2-3}(0) a_{2}\;,
\label{eq:r123} \\
\rho_{2-3}^{(0)} & \simeq & \rho_{2-3}(0)\;,
\label{eq:r023}
\end{eqnarray}
\end{mathletters}
which yield, respectively,
\begin{mathletters}
\label{eq:tr23r23}
\begin{eqnarray}
{\rm Tr}_{2-3}\rho_{2-3}^{(1)} & = & 
\chi_{1}^{2}t^{2}\left(\bar{N}+1\right)\;,
\label{eq:tr23r123} \\
{\rm Tr}_{2-3} \rho_{2-3}^{(0)} & \simeq & 1\;,
\label{eq:tr23r023}
\end{eqnarray}
\end{mathletters}
and, finally,
\begin{equation}
e^{i\phi}\langle a_{2o}^{\dagger} a_{2e}\rangle_{\rm o-d}
= \frac{\bar{N}+1}{2} e^{i\phi}\;.
\end{equation}
In conclusion, considering the off-diagonal contribution, 
Eq.~(\ref{eq:cdccond}) can be rewritten as
\begin{equation}
\langle  c^{\dagger}c\rangle_{\rm cat} = 
\langle  c^{\dagger}c\rangle_{\rm mix}
+ \frac{\bar{N}+1}{2} \cos\phi\;.
\label{eq:cdccondmix}
\end{equation}
It is then clear that the photocounts at the detector D$_{\rm c}$ 
exhibit interference fringes as a function of the variable phase 
$\phi$, if and only if the state~(\ref{eq:2erhored}) is a true linear 
superposition and not just a statistical mixture of the two 
macroscopic components. The visibility of such interference fringes
is given by
\begin{equation}
V=\frac{1+\bar{N}}{1+3\bar{N}}
\label{eq:visi}
\end{equation}
and has therefore the lower bound $1/3$ for $\bar{N}\to \infty$.

\subsection{Correlation functions}
\label{corr}

Our aim in this subsection is to compute the first- and second-order
correlation functions relative to our output modes, in order to make
an independent test of the presence of quantum coherence in our system.
We keep in mind~\cite{kn:miwa} that a manifestation of quantum 
coherence at second order is subpoissonian statistics, i.e.,
\begin{equation}
G^{(2)}(0) < \left[G^{(1)}(0)\right]^{2}\;,
\label{eq:subpo}
\end{equation}
where $G^{(1)}(0)$ and $G^{(2)}(0)$ are, respectively, the first-
and second-order correlation functions.

Let us consider the same experimental apparatus we have proposed
for the detection of interference (see Fig.~\ref{fg:scheme}). We
take now into account both output ports $c$ and $d$ of the polarizing
beam splitter, with annihilation operators
\begin{mathletters}
\label{eq:cd}
\begin{eqnarray}
c=\frac{1}{\protect\sqrt{2}} \left( a_{2o} +e^{i\phi} a_{2e} 
\right)\;,
\label{eq:cc} \\
d=\frac{1}{\protect\sqrt{2}} \left( a_{2o} -e^{i\phi} a_{2e} 
\right)\;,
\label{eq:dd}
\end{eqnarray}
\end{mathletters}
and evaluate the correlation functions
$\langle c^{\dagger}cc^{\dagger}c \rangle$,
$\langle d^{\dagger}dd^{\dagger}d \rangle$,
and $\langle c^{\dagger}cd^{\dagger}d \rangle$,
where $c^{\dagger}c$ is given by Eq.~(\ref{eq:nc}), and
\begin{equation}
d^{\dagger}d=\frac{1}{2}\left(
a_{2o}^{\dagger}a_{2o}
+a_{2e}^{\dagger}a_{2e}
-e^{i\phi}a_{2o}^{\dagger}a_{2e}
-e^{-i\phi}a_{2e}^{\dagger}a_{2o}\right)\;.
\label{eq:nd}
\end{equation}

We shall evaluate the functions $(c^{\dagger}c)^{2}$, 
$(d^{\dagger}d)^{2}$, and $c^{\dagger}c d^{\dagger}d$ in the 
small-time approximation limit (see appendix), in which
\begin{eqnarray}
\rho_{2e-3o-2o-3e} (t) & \propto &
\left(a_{2e}^{\dagger}+a_{2o}\dagger\right)
\rho_{2e-3o} (0) \rho_{2o-3e} (0)
\nonumber \\
 & & \times \Big( a_{2e}+a_{2o} \Big)\;, 
\label{eq:r23st}
\end{eqnarray}
where $\rho_{2o-3e} (0)$ is the Gaussian state described by the Wigner
function $W_{\rm bt}^{2-3} (0)$ of Eq.~(\ref{eq:w23bt}), for which
the Wigner function corresponding to the reduced density matrix of
mode $2$ alone is given by Eq.~(\ref{eq:redtherm}),
that represents a thermal state with a mean number of photons given by
$\bar{N}$ of Eq.~(\ref{eq:n20}).

Upon evaluating all the required expectation values, we obtain
\begin{mathletters}
\label{eq:corr}
\begin{eqnarray}
\langle \left(c^{\dagger}c\right)^{2}\rangle & = &
\frac{16\bar{N}^{2}+14\bar{N} + 2 + 2\cos\phi
\left(4\bar{N}^{2}+5\bar{N}+1\right)}{4}
\nonumber \\
 & & \label{eq:corrc} \\
\langle \left(d^{\dagger}d\right)^{2}\rangle & = &
\frac{16\bar{N}^{2}+14\bar{N} + 2 - 2\cos\phi
\left(4\bar{N}^{2}+5\bar{N}+1\right)}{4}
\nonumber \\
 & & \label{eq:corrd} \\
\langle c^{\dagger}cd^{\dagger}d\rangle & = &
\bar{N}\left(2\bar{N}+1\right)\;.
\label{eq:corrcd}
\end{eqnarray}
\end{mathletters}
From Eqs.~(\ref{eq:corr}) it is clear that the visibility of the 
fringes in $\langle (c^{\dagger}c)^{2}\rangle$ and
$\langle (d^{\dagger}d)^{2}\rangle$ is given by 
\begin{equation}
V = \frac{4\bar{N}^{2}+5\bar{N} + 1}{8\bar{N}^{2} + 7\bar{N} + 1}\;,
\label{eq:corrvis}
\end{equation}
and monotonically decreases from $V=1$ (for $\bar{N}=0$) to $V=1/2$
(for $\bar{N} \to \infty$).

Finally, considering the field at the output port $c$, the first-
and second-order correlation functions for mode $2$ can be written as
\begin{mathletters}
\label{eq:gg}
\begin{eqnarray}
G^{(1)}_{2}(0) & = & \langle c^{\dagger}c \rangle =
\frac{1+3\bar{N} + (1+\bar{N})\cos\phi}{2}\;,
\label{eq:gg1} \\
G^{(2)}_{2}(0) & = & \langle c^{\dagger}c^{\dagger}cc\rangle =
\langle \left(c^{\dagger}c\right)^{2}\rangle - \langle c^{\dagger}c
\rangle
\nonumber \\
 & = & 2\bar{N}\left[1+2\bar{N}+\left(\bar{N}+1\right)\cos\phi\right]\;,
\label{eq:gg2}
\end{eqnarray}
\end{mathletters} 
respectively. It should be noted that these results map into the 
corresponding ones obtained in Ref.~\cite{kn:dem} for the OPA case 
upon a redefinition of the phase angles.
By comparing $[G^{(1)}(0)]^{2}$ and $G^{(2)}(0)$ it is
possible to see that $G^{(2)}(0) < [G^{(1)} (0)]^{2}$ only at low
mean photon number, as it could have been easily expected.
The best situation is obtained when $\phi =0$, in which case
\begin{mathletters}
\label{eq:ggg}
\begin{eqnarray}
[G^{(1)}(0)]^{2} & = & \left(1+2\bar{N}\right)^{2}\;,
\label{eq:ggg1} \\
G^{(2)}(0) & = & 2\bar{N}\left(3\bar{N}+2\right)\;,
\label{eq:ggg2}
\end{eqnarray}
\end{mathletters}
and the condition for quantum coherence at second order is reached 
when $\bar{N} < 1/\sqrt{2}$.
On the other hand, when $\phi=\pi$, $[G^{(1)}(0)]^{2}=\bar{N}^{2}$,
$G^{(2)}(0)=2\bar{N}^{2}$, and therefore $G^{(2)}(0)$ is always larger
than $[G^{(1)}(0)]^{2}$.

\subsection{Wigner function}
\label{wigner}

The aim of the present section is to provide a means to represent the 
essential features of the Schr\"odinger-cat state, Eq.~(\ref{eq:rhocon}),
which ``lives'' in a 8-dimensional phase space, in the more customary
2-dimensional phase space, in order to make a comparison with the more
conventional cat states~\cite{kn:cat,kn:prlha}.
Let us start from Eq.~(\ref{eq:rhocon}) which we rewrite here for 
convenience
\begin{eqnarray}
\rho^{c}_{2o-3e-2e-3o}(t) & \propto & 
\left[\rho^{(1)}_{2e-3o}\rho^{(0)}_{2o-3e}
+\rho^{(0)}_{2e-3o}\rho^{(1)}_{2o-3e}\right.
\nonumber \\
 + & & \left.\rho^{({\rm int})}_{2e-3o}\rho^{({\rm int})\,\dagger}_{2o-3e}
+\rho^{({\rm int})\,\dagger}_{2e-3o}\rho^{({\rm 
int})}_{2o-3e}\right]\;.
\label{eq:cat}
\end{eqnarray}
The Wigner function representation of the density matrix~(\ref{eq:cat})
would of course reflect its characteristic Schr\"odinger-cat properties.
However, in order to better understand the nature of this state, it 
would be interesting and desirable to see whether it is possible to 
find different optical modes in whose terms the state (and therefore
the Wigner function) may be rewritten in a simpler form. Our key idea is 
then to look for linear combinations of mode operators (which can
easily be realized with linear elements: polarizers and beam-splitters)
such as to factorize the state~(\ref{eq:cat}) in smaller subspaces.

We first perform a transformation which changes the horizontally and
vertically polarized modes into the 45$^{\circ}$-polarized ones, 
namely,
\begin{mathletters}
\label{eq:modetra}
\begin{eqnarray}
a_{+45,2} & = & \frac{a_{2e}+a_{2o}}{\protect\sqrt{2}}\;,
\;\;\;
a_{-45,2}=\frac{a_{2e}-a_{2o}}{\protect\sqrt{2}}\;,
\label{eq:amodetra} \\
a_{+45,3} & = & \frac{a_{3e}+a_{3o}}{\protect\sqrt{2}}\;,
\;\;\;
a_{-45,3}=\frac{a_{3e}-a_{3o}}{\protect\sqrt{2}}\;,
\label{eq:bmodetra}
\end{eqnarray}
\end{mathletters}
and the corresponding expressions for mode $\vec{k}_{1}$ and for
the creation operators. In terms 
of these new operators, $H_{\rm NL1}$ and $H_{\rm NL2}$ 
[Eqs.~(\ref{eq:hnl1}) and (\ref{eq:hnl2})] can be rewritten as
\begin{mathletters}
\label{eq:hnlp}
\begin{eqnarray}
H_{\rm NL1} & = & i\hbar\chi_{1}\left({a}_{+45,2}^{\dagger}
{a}_{+45,1}^{\dagger}-{a}_{+45,2}{a}_{+45,1}\right)
\nonumber \\
& & -i\hbar\chi_{1}
\left({a}_{-45,2}^{\dagger}{a}_{-45,1}^{\dagger}
-{a}_{-45,1}{a}_{-45,2}\right)\;,
\label{eq:hnl1p} \\
{H}_{\rm NL2} & = & i\hbar\chi_{2}\left({a}_{+45,2}^{\dagger}
{a}_{+45,3}^{\dagger}-{a}_{+45,2}{a}_{+45,3}\right)
\nonumber \\
& & -i\hbar\chi_{2}
\left({a}_{-45,2}^{\dagger}{a}_{-45,3}^{\dagger}
-{a}_{-45,2}{a}_{-45,3}\right)
\label{eq:hnl2p}\;.
\end{eqnarray}
\end{mathletters}
We have already assumed [Sec.~\ref{evol}] that the cavity decay rates
$\kappa_{i}$ do not depend on the polarization. This in turn means that
$\kappa_{+45,2}=\kappa_{-45,2}=\kappa_{2}$ and
$\kappa_{+45,3}=\kappa_{-45,3}=\kappa_{3}$, and therefore we have that
for the $\pm 45^{\circ}$-polarized modes we have the same evolution equation as 
that for the original modes (except for a minus sign).
Consequently, it is possible to repeat all the same arguments as before
[Secs.~\ref{evol} and \ref{genera}]. In particular, the modes
$a_{+45,1}$, $a_{+45,2}$, and $a_{+45,3}$ are decoupled from their
orthogonal counterparts $a_{-45,1}$, $a_{-45,2}$, and $a_{-45,3}$,
and the evolution equation may be rewritten as
\begin{equation}
\rho_{\rm T}(t)=e^{{\cal L}_{+45}t} e^{{\cal L}_{-45}t} 
\rho_{+45,1;+45,2;+45,3;-45,1;-45,2;-45,3} (0)\;.
\label{eq:evolrt}
\end{equation}
In Eq.~(\ref{eq:evolrt}) the initial condition is given in the same 
way by
\begin{equation}
\rho_{+45,1;+45,2;+45,3}(0) = |0\rangle_{+45,1}\langle 0|
\otimes \rho^{\rm bt}_{+45,2;+45,3} (0) \;,
\label{eq:in45}
\end{equation}
where $\rho^{\rm bt}_{+45,2;+45,3} (0)$ is the equilibrium state below
threshold of the parametric oscillator when NL1 is turned off, and the
same initial condition holds for the $-45^{\circ}$-polarized modes.
As a consequence, the same Gaussian evolution we have found in 
Sec.~\ref{evol} holds. The only difference is that now the conditional 
measurement is simply a projection onto the state $|1\rangle_{+45,1}$,
i.e., the one-photon state for the $a_{+45,1}$ mode, while the
$-45^{\circ}$-polarized modes remain decoupled from the orthogonal ones.

The cat state after the conditional detection of the $n=1$ photon for 
the $+45,1$ mode is then written in the following way

\end{multicols}
\vspace{-0.6cm}
\noindent\rule{0.5\textwidth}{0.4pt}\rule{0.4pt}{\baselineskip}
\widetext

\begin{eqnarray}
\rho^{\rm c} & \propto & {}_{-45,1}\langle 0|_{+45,1}\langle 1 |
\rho_{+45,1;+45,2;+45,3} (t)
\rho_{-45,1;-45,2;-45,3} |1\rangle_{+45,1} 
|0\rangle_{-45,1} =
\rho_{+45,2;+45,3}^{(1)} \otimes \rho_{-45,2;-45,3}^{(0)} \;, 
\label{eq:cr01}
\end{eqnarray}

\noindent\hfill\rule{0.5\textwidth}{0.4pt}

\noindent\null\hfill\rule{0.4pt}{\baselineskip}\hfill\null

\begin{multicols}{2}
\vspace{-0.5cm}
\noindent
where $\rho_{2-3}^{(0)}$ and $\rho_{2-3}^{(1)}$ are again given by the
expressions~(\ref{eq:rhocon231}) and (\ref{eq:rhocon230}).
It should be noted that, using these new $\pm 45^{\circ}$-polarized 
modes, one gets a complete factorization of the $-45^{\circ}$-polarized
modes, which are {\em not affected} by the {\em quantum injection}
process induced by the conditional measurement. The 
$-45^{\circ}$-polarized modes are not ``interesting'', in the sense 
that all the ``cat'' properties of the state~(\ref{eq:cr01}) are 
contained in $\rho_{+45,2;+45,3}^{(1)}$, and therefore we shall 
neglect them from now on. We are then left with the state
$\rho_{+45,2;+45,3}^{(1)}$, which is an {\em entangled} state of the
modes $+45,2$ and $+45,3$.

\begin{figure}
\centerline{\psfig{figure=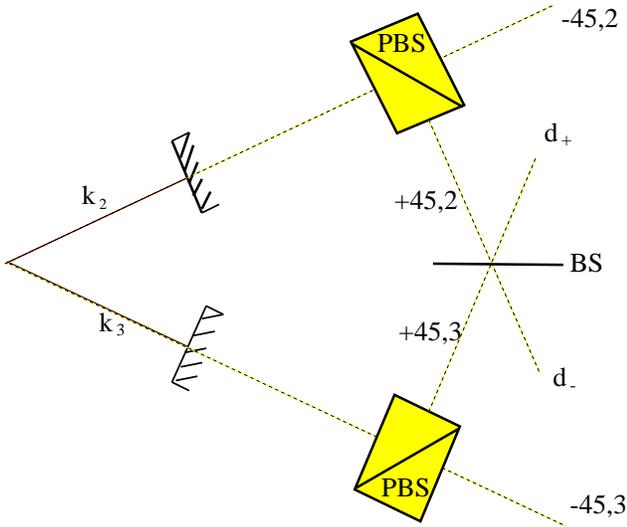,width=3.3in}}
\vspace{0.2cm}
\caption{\narrowtext
         Scheme of the experimental arrangement needed for the 
         measurement of the Wigner function (see text). $k_2$ and $k_3$
         represent the cavity modes of Fig.~\protect\ref{fg:scheme}}
\label{fg:schemwig}
\end{figure}

As the second step of our procedure aimed at the further 
simplification of the original 8-dimensional Wigner function, we 
consider the transformation
\begin{equation}
d_{+} = \frac{a_{+45,2}+a_{+45,3}}{\protect\sqrt{2}}\;,
\;\;\;
d_{-} = \frac{a_{+45,2}-a_{+45,3}}{\protect\sqrt{2}}\;,
\label{eq:d+d-}
\end{equation}
which is suggested by the interaction term in Eq.~(\ref{eq:hnl2p}).
In terms of $d_{+}$ and $d_{-}$, Eq.~(\ref{eq:hnl2p}) becomes
\begin{equation}
H_{\rm NL2} = i\hbar\frac{\chi_{2}}{2}\left(d_{+}^{\dagger \, 2}
-d_{+}^{2}\right) -i\hbar\frac{\chi_{2}}{2}\left(d_{-}^{\dagger \, 2}
- d_{-}^{2}\right)\;,
\label{eq:hnl2d}
\end{equation}
and the two modes $d_{+}$ and $d_{-}$ are squeezed by the nonlinear 
crystal. These modes can be experimentally realized outside the cavity
for example with two PBS and a 50\%--50\% BS, as schematically described in 
Fig.~\ref{fg:schemwig}. The state of these two modes can be 
represented by the Wigner function
\begin{equation}
W_{1} \left(
\frac{x_{d_{+}}+x_{d_{-}}}{\protect\sqrt{2}},
\frac{y_{d_{+}}+y_{d_{-}}}{\protect\sqrt{2}},
\frac{x_{d_{+}}-x_{d_{-}}}{\protect\sqrt{2}},
\frac{y_{d_{+}}-y_{d_{-}}}{\protect\sqrt{2}}
\right)\;,
\label{eq:w1d+d-}
\end{equation}
where [see Eqs.~(\ref{eq:sol123}) and (\ref{eq:bt})]

\end{multicols}
\vspace{-0.6cm}
\noindent\rule{0.5\textwidth}{0.4pt}\rule{0.4pt}{\baselineskip}
\widetext

\begin{eqnarray}
W_{1} \left( x_{2},y_{2},x_{3},y_{3} \right)
 & \propto & \int dx_{1} dy_{1} \, W_{n=1} \left( x_{1}, y_{1} \right)
\times \frac{\protect\sqrt{\det B}}{\pi^{2}} e^{-\langle \vec{z} |
 B | \vec{z} \rangle}\;.
\label{eq:w12233}
\end{eqnarray}

What is the nature of this state? In order to answer this question,
we are naturally guided by two different approaches: i) the study
of the OPA case~\cite{kn:dem} and ii) the use of the small-time limit
$\chi_{1} t,\chi_{2} t,\kappa t \ll 1$ we have already considered in
Sec.~\ref{photo} and worked out in the appendix.
In the OPA case~\cite{kn:dem} the output state at time $t$ is given by
\begin{eqnarray}
|\psi(t)\rangle & = & \frac{1}{\protect\sqrt{2}}
 e^{\chi_{2}t (a_{2e}^{\dagger}a_{3o}^{\dagger} - a_{2e} a_{3o})
 + \chi_{2}t(a_{2o}^{\dagger}a_{3e}^{\dagger} - a_{2o} a_{3e})}
\times \left(a_{2e}^{\dagger}+a_{2o}^{\dagger} \right) |0\rangle\;,
\label{eq:opapsi}
\end{eqnarray}
which can be rewritten in terms of the $\pm 45^{\circ}$-polarized 
modes as
\begin{eqnarray}
|\psi (t) \rangle & = & 
e^{-\chi_{2}t(a_{-45,2}^{\dagger}a_{-45,3}^{\dagger}-a_{-45,2}a_{-45,3})}
|0\rangle
\otimes e^{\chi_{2}t(a_{+45,2}^{\dagger}a_{+45,3}^{\dagger}
 -a_{+45,2}a_{+45,3})} a_{+45,2}^{\dagger} |0\rangle=
\psi_{-45,2;-45,3}^{(0)} \psi_{+45,2;+45,3}^{(1)}\;.
\label{eq:opapsi45}
\end{eqnarray}
Neglecting the factorized state $\psi_{-45,2;-45,3}^{(0)}$, and using
the $d_{\pm}$ modes, we have
\begin{mathletters}
\label{eq:squent}
\begin{eqnarray}
\psi_{+45,2;+45,3}^{(1)} & = &
e^{\frac{\chi_{2}t}{2}(d_{+}^{\dagger\, 2}-d_{+}^{2})}
e^{-\frac{\chi_{2}t}{2}(d_{-}^{\dagger\, 2}-d_{-}^{2})}
 \left(\frac{d_{+}^{\dagger}+d_{-}^{\dagger}}{\protect\sqrt{2}}\right)
|0\rangle
\label{eq:squenta} \\
 & = &
\frac{1}{\protect\sqrt{2}}
 \left( \left|\frac{\chi_{2}t}{2}, 1\right\rangle_{d_{+}}
 \left|-\frac{\chi_{2}t}{2}, 0\right\rangle_{d_{-}}\right.
+ \left.\left|\frac{\chi_{2}t}{2}, 0\right\rangle_{d_{+}}
 \left|-\frac{\chi_{2}t}{2}, 1\right\rangle_{d_{-}} \right)\;,
\label{eq:squentb}
\end{eqnarray}
\end{mathletters}
\noindent\hfill\rule{0.5\textwidth}{0.4pt}

\noindent\null\hfill\rule{0.4pt}{\baselineskip}\hfill\null

\begin{multicols}{2}
\vspace{-0.6cm}
\noindent
which is an entangled superposition of the {\em squeezed} one-photon
and vacuum states of the modes $d_{+}$ and $d_{-}$.
It is quite clear now that if we want to ``isolate'' one mode, say,
the $d_{+}$ mode, we need a second {\em conditional measurement} on
the mode $d_{-}$, e.g., a projection onto the state
\begin{equation}
|\varphi\rangle_{d_{-}} = \alpha |0\rangle_{d_{-}}
+ \beta |1\rangle_{d_{-}}\;.
\label{eq:prosup}
\end{equation}
Such a conditional measurement could be performed, for example, by sending a 
two-level atom---resonant with the atomic transition---through the 
cavity, and eventually post-selecting its internal state in a 
corresponding superposition of its ground and excited states.
The conditional state, provided the measurement has given a successful
result, would then read as
\begin{equation}
|\psi^{\rm c}\rangle_{d_{+}} \propto \alpha^{\star} |\frac{\chi_{2}t}{2}
, 1\rangle_{d_{+}} + \frac{\beta^{\star}}{\cosh \chi_{2}t}
|\frac{\chi_{2}t}{2},0\rangle_{d_{+}}\;.
\label{eq:psicd+}
\end{equation}

We can reach a similar conclusion also by analyzing the OPO case using 
the very well justified small-time approximation (see appendix) in the
limit $\chi_{1}t, \chi_{2}t, \kappa t \ll 1$, applied to the modes
$+45,1$, $+45,2$, and $+45,3$.
We have, at the lowest order in $\chi_{1}t$,
\begin{mathletters}
\label{eq:r12345}
\begin{eqnarray}
\rho_{+45,2;+45,3}^{(1)} & = & {}_{+45,1}\langle 1|\rho_{1-2-3} 
|1\rangle_{+45,1}
\label{eq:r12345a} \\
 & \propto & a_{+45,2}^{\dagger} \rho_{2-3}(0) a_{+45,2}\;,
\label{eq:r12345b}
\end{eqnarray}
\end{mathletters}
where the initial density matrix $\rho_{2-3}(0)$ is the state described
by the Wigner function~(\ref{eq:w23bt}).
If we now write Eq.~(\ref{eq:w23bt}) in terms of the new variables
corresponding to the modes $d_{+}$ and $d_{-}$, namely,
\begin{mathletters}
\label{eq:varchange}
\begin{eqnarray}
x_{d_{+}} & = & \frac{x_{2}+x_{3}}{\protect\sqrt{2}}\;, \;\;\;\;\;\;
x_{d_{-}} = \frac{x_{2}-x_{3}}{\protect\sqrt{2}}\;,
\label{eq:varchana} \\
y_{d_{+}} & = & \frac{y_{2}+y_{3}}{\protect\sqrt{2}}\;, \;\;\;\;\;\;
x_{d_{-}} = \frac{y_{2}-y_{3}}{\protect\sqrt{2}}\;,
\label{eq:varchanb}
\end{eqnarray}
\end{mathletters}
we obtain

\end{multicols}
\vspace{-0.6cm}
\noindent\rule{0.5\textwidth}{0.4pt}\rule{0.4pt}{\baselineskip}
\widetext

\begin{eqnarray}
W_{\rm bt}^{d_{+}d_{-}} & = & W_{\rm bt}^{d_{+}} W_{\rm bt}^{d_{-}} =
\frac{2}{\pi} \sqrt{1-\frac{\chi_{2}^{2}}{\kappa^{2}}}
 \exp\left[-2\left(1-\frac{\chi_{2}}{\kappa}\right) x_{d_{+}}^{2}\right.
      \left.-2\left(1+\frac{\chi_{2}}{\kappa}\right) y_{d_{+}}^{2}\right]
\nonumber \\
 & & \times
\frac{2}{\pi} \sqrt{1-\frac{\chi_{2}^{2}}{\kappa^{2}}}
 \exp\left[-2\left(1+\frac{\chi_{2}}{\kappa}\right) x_{d_{-}}^{2}\right.
 \left. -2\left(1+\frac{\chi_{2}}{\kappa}\right) y_{d_{-}}^{2}\right]\;.
\nonumber \\
\label{eq:wdb}
\end{eqnarray}
The initial states for the modes $d_{\pm}$ are generalized Gaussian
states~\cite{kn:gar}, of the kind
\begin{equation}
\rho_{\pm} \propto \exp \left( -nd_{\pm}^{\dagger}d_{\pm}
-\frac{1}{2} m^{\star}_{\pm} d_{\pm}^{2}-\frac{1}{2} m_{\pm} 
d_{\pm}^{\dagger \, 2} \right)\;,
\label{eq:gengau}
\end{equation}
with
\begin{mathletters}
\label{eq:nmgau}
\begin{equation}
n=\frac{1}{\protect\sqrt{1-\frac{\chi_{2}^{2}}{\kappa^{2}}}} \log
\left(\frac{1+\protect\sqrt{1-\frac{\chi_{2}^{2}}{\kappa^{2}}}}
{1-\protect\sqrt{1-\frac{\chi_{2}^{2}}{\kappa^{2}}}} \right)\;,
\label{eq:ngau}
\end{equation}
and
\begin{equation}
m_{\pm} = \pm \frac{\chi_{2}}{\kappa} n\;.
\label{eq:mgau}
\end{equation}
\end{mathletters}
Since the initial state factorizes, we have
\begin{mathletters}
\label{eq:rfac}
\begin{eqnarray}
\rho_{d_{+}d_{-}}^{(1)} & \propto & a_{+45,2}^{\dagger} \rho_{d_{+}}(0)
\rho_{d_{-}} (0) a_{+45,2}
\propto (d_{+}^{\dagger}+d_{-}^{\dagger}) \rho_{d_{+}}(0)
\rho_{d_{-}} (0) (d_{+}+d_{-})
\label{eq:rfacb} \\
 & = & \left(d_{+}^{\dagger} \rho_{d_{+}} (0) d_{+} \right) \otimes 
 \rho_{d_{-}} (0)
+ \rho_{d_{+}} (0) \otimes \left(d_{-}^{\dagger}
 \rho_{d_{-}} (0) d_{-}^{\dagger} \right)
\nonumber \\
 & &
+ \left(d_{+}^{\dagger} \rho_{d_{+}} (0) \right) \otimes
 \left(\rho_{d_{-}} (0) d_{-} \right)
+ \left(\rho_{d_{+}} (0) d_{+} 
 \right) \otimes \left( d_{-}^{\dagger} \rho_{d_{-}} (0) \right)\;,
\label{eq:rfacc}
\end{eqnarray}
\end{mathletters}
which is a mixed state analogous to the pure state~(\ref{eq:squent})
obtained in the OPA case. Its Wigner function can be calculated from
Eq.~(\ref{eq:rfac}) and is given by
\begin{eqnarray}
 W\left(x_{d_{+}}, y_{d_{+}}, x_{d_{-}}, y_{d_{-}}\right) & = & 
\frac{4}{\pi^{2}} \frac{\left(1-\frac{\chi_{2}^{2}}{\kappa^{2}}
\right)^{2}}{2-\frac{\chi_{2}^{2}}{\kappa^{2}}}
\left[\left(x_{d_{+}}^{2}+y_{d_{-}}^{2}\right)
\left(2-\frac{\chi_{2}}{\kappa}\right)^{2} +
\left(y_{d_{+}}^{2}+x_{d_{-}}^{2}\right)
\left(2+\frac{\chi_{2}}{\kappa}\right)^{2} -2\right.
\nonumber \\
 & & \;\; \left.+2\left(4-\frac{\chi_{2}^{2}}{\kappa^{2}}\right)
 \left(x_{d_{+}}x_{d_{-}}+y_{d_{+}}y_{d_{-}}\right)\right]
 e^{-2\left(1-\frac{\chi_{2}}{\kappa}\right)
 \left(x_{d_{+}}^{2}+y_{d_{-}}^{2}\right)
 -2\left(1+\frac{\chi_{2}}{\kappa}\right)
 \left(x_{d_{-}}^{2}+y_{d_{+}}^{2}\right)}\;.
\label{eq:wigfour}
\end{eqnarray}

\noindent\hfill\rule{0.5\textwidth}{0.4pt}

\noindent\null\hfill\rule{0.4pt}{\baselineskip}\hfill\null

\begin{multicols}{2}
\vspace{-0.6cm}
\noindent
Two important features should be noted within the form of this Wigner 
function: i) the interference term (the last term in the square
brackets) decreases when the number of photons in the initial state
increases. This behavior is governed by the factor
$4-\chi_{2}^{2}/\kappa^{2}$ and by the fact that [see 
Eqs.~(\ref{eq:meannumb}) and (\ref{eq:n2})] $\bar{N} \to \infty$
when $\chi_{2}/\kappa \to 1$. ii) The Wigner function is negative
around the origin and its negativity scales to zero as the
initial mean photon number $\bar{N} \to \infty$. In fact,
\begin{equation}
W(0,0,0,0)=-\frac{8}{\pi^{2}}\frac{\left(1-\frac{\chi_{2}^{2}}{\kappa^{2}}
\right)^{2}}{2-\frac{\chi_{2}^{2}}{\kappa^{2}}} =
-\frac{4}{\pi^{2}}\frac{1}{(2\bar{N}+1)(\bar{N}+1)}\;.
\label{eq:wzero}
\end{equation}
We have already seen the same scaling behavior of quantum properties
with $\bar{N} \to \infty$ in the calculation of the second-order
correlation function $G^{(2)}$: This is one of the desired properties
of a Schr\"odinger-cat state, as we have emphasized at the beginning
of this section.

Again, Eq.~(\ref{eq:wigfour}) bears a remarkable similarity with the 
corresponding result obtained in Ref.~\cite{kn:dem} for the OPA 
configuration, in the limits $\kappa \to \infty$ and of small 
interaction times. The main advantage of the OPO is given by the 
larger effective number of photons per mode $N$
[see Eq.~(\ref{eq:meannumb})] with respect to the
$\sinh^{2}\chi t$ of the OPA~\cite{kn:dem}.

We have therefore learnt that in order to obtain a one-mode state
which embodies all the relevant features of the original four-mode
cat state one {\em has to perform} a conditional measurement on the
mode $d_{-}$. When this is successfully done, the final conditioned
state of the mode $d_{+}$ alone is described by the Wigner function
\begin{equation}
W(\delta_{+}) \propto \pi \int d^{2} \delta_{-}
W_{\alpha |0\rangle + \beta |1\rangle} (\delta_{-})
W(\delta_{+},\delta_{-})\;,
\label{eq:wigd+}
\end{equation}
where
\begin{eqnarray}
 & & W(\delta_{+},\delta_{-}) =
\nonumber \\
& & W_{1} \left(
\frac{x_{d_{+}}+x_{d_{-}}}{\protect\sqrt{2}},
\frac{y_{d_{+}}+y_{d_{-}}}{\protect\sqrt{2}},
\frac{x_{d_{+}}-x_{d_{-}}}{\protect\sqrt{2}},
\frac{y_{d_{+}}-y_{d_{-}}}{\protect\sqrt{2}}
\right)\;,
\nonumber \\ 
\label{eq:wde+-}
\end{eqnarray}
is the Wigner function [see Eqs.~(\ref{eq:w1d+d-}) and
(\ref{eq:w12233})] of the state~(\ref{eq:rfac}), and
\begin{eqnarray}
W_{\alpha |0\rangle + \beta |1\rangle} (\delta_{-})
 & = & \frac{2}{\pi} 
\Big[1+4|\beta|^{2}\left(|\delta|^{2}-\frac{1}{2}\right)
\nonumber \\
 & & +4{\rm Re} (\delta){\rm Re}(\alpha\beta^{\star})
 +4{\rm Im}(\delta){\rm Im} (\alpha^{\star}\beta) \Big]
 \nonumber \\
\label{eq:wab}
\end{eqnarray}
is the Wigner function of the state onto which the conditional
measurement projects the mode $d_{-}$ [Eq.~(\ref{eq:prosup})].
According to the small-time limit approximation (see appendix)
the explicit form of the Wigner function~(\ref{eq:wigd+}) can
be derived from Eqs.~(\ref{eq:rfac})--(\ref{eq:wab}) and,
after a lengthy calculation, reads as

\end{multicols}
\vspace{-0.6cm}
\noindent\rule{0.5\textwidth}{0.4pt}\rule{0.4pt}{\baselineskip}
\widetext

\begin{eqnarray}
W(x_{d_{+}},y_{d_{+}}) & \propto &
\exp\left[-2\left(1-\frac{\chi_{2}}{\kappa}\right)x_{d_{+}}^{2}
-2\left(1+\frac{\chi_{2}}{\kappa}\right)y_{d_{+}}^{2}\right]
\Bigg\{\left(|\alpha|^{2}
  +|\beta|^{2}\frac{\chi_{2}^{2}/\kappa^{2}}{4-\chi_{2}^{2}/\kappa^{2}}\right)
\left[x_{d_{+}}^{2}\left(2-\frac{\chi_{2}}{\kappa}\right)^{2}\right.
\nonumber \\
 & & \left.+ y_{d_{+}}^{2}\left(2+\frac{\chi_{2}}{\kappa}\right)^{2} -1\right]
+ |\beta|^{2} + 2{\rm Re} \left(\alpha^{\star}\beta
 \left[x_{d_{+}}\left(2-\frac{\chi_{2}}{\kappa}\right)\right.\right.
-iy_{d_{+}}
 \left.\left.\left(2+\frac{\chi_{2}}{\kappa}\right)\right]\right)\Bigg\}\;,
\label{eq:wd}
\end{eqnarray}

\noindent\hfill\rule{0.5\textwidth}{0.4pt}

\noindent\null\hfill\rule{0.4pt}{\baselineskip}\hfill\null

\begin{figure}
\centerline{\psfig{figure=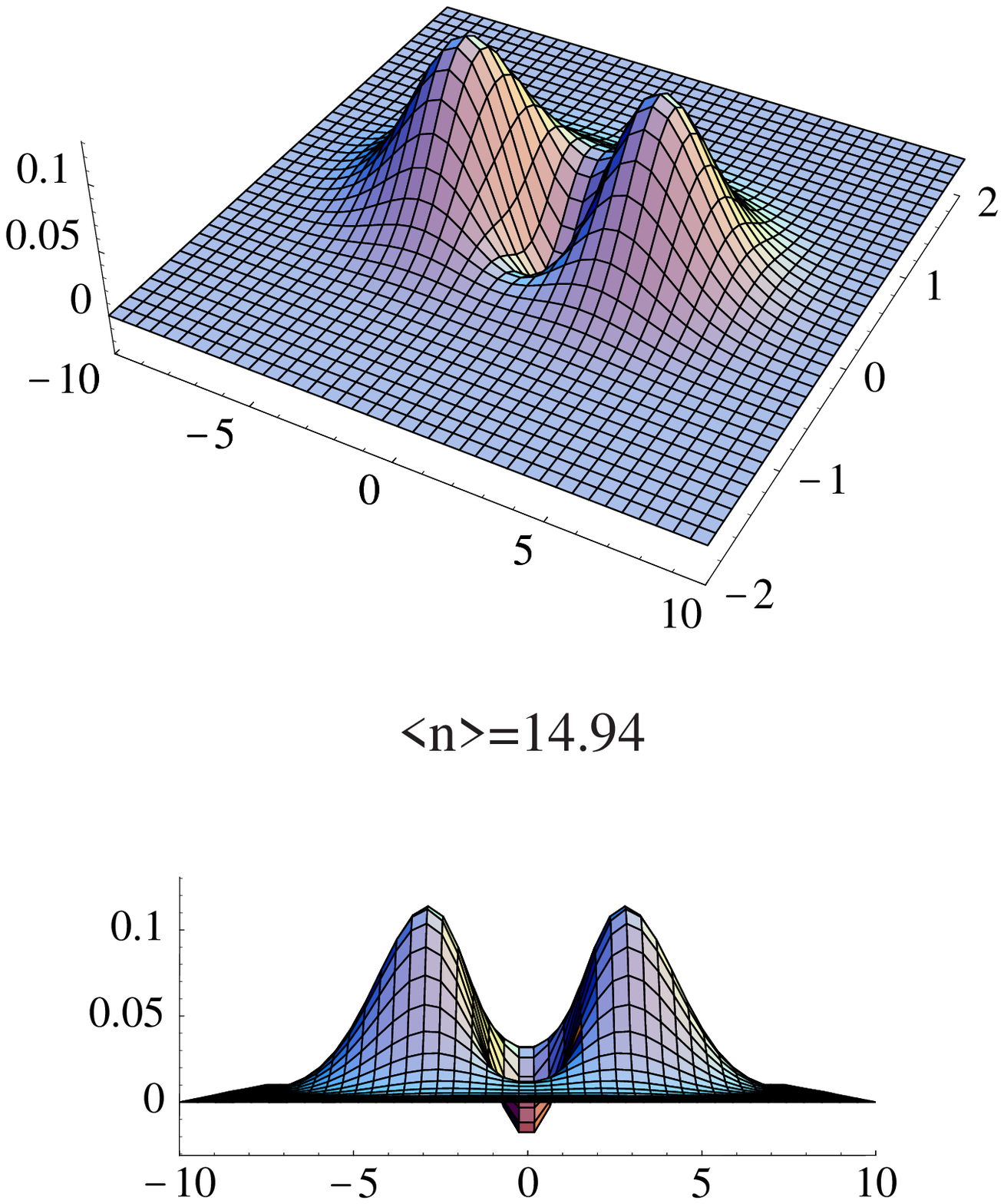,width=3.3in}}
\vspace{0.1cm}
\caption{\narrowtext
         Tridimensional plot of the Wigner function for the $d_{+}$ mode
         for an initial mean number of photons $\langle n \rangle=N=14.94$.
         Two different viewpoints have 
         been chosen to display more clearly the quantum 
         superposition character of the Schr\"odinger-cat state.}
\label{fg:wig1}
\end{figure}

\begin{figure}
\centerline{\psfig{figure=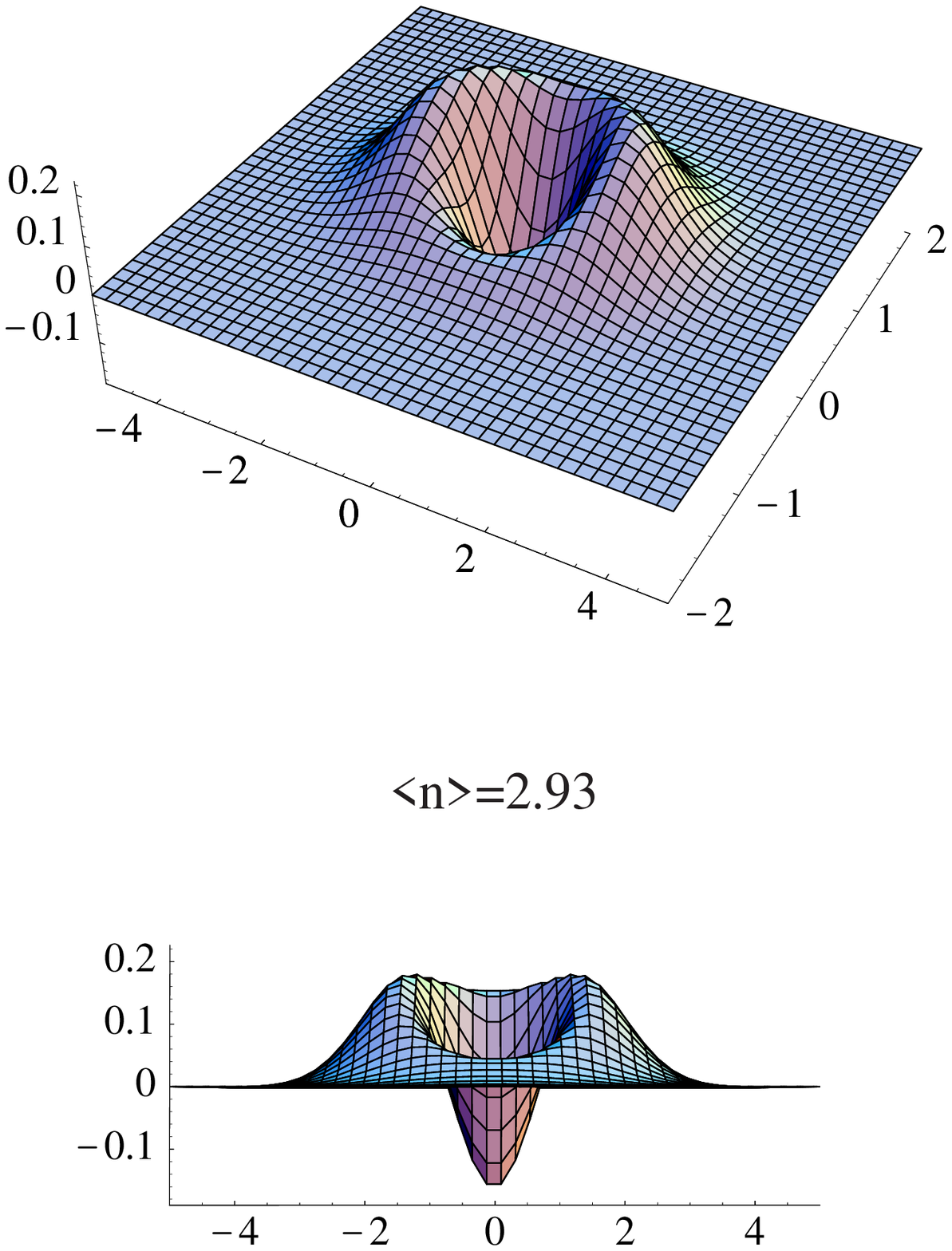,width=3.3in}}
\vspace{0.1cm}
\caption{\narrowtext
         Tridimensional plot of the Wigner function for the $d_{+}$ mode
         for an initial mean number of photons $\langle n \rangle=N=2.93$.
         In (a) and (b) two different viewpoints have 
         been chosen to display more clearly the quantum 
         superposition character of the Schr\"odinger-cat state.}
\label{fg:wig2}
\end{figure}

\begin{figure}
\centerline{\psfig{figure=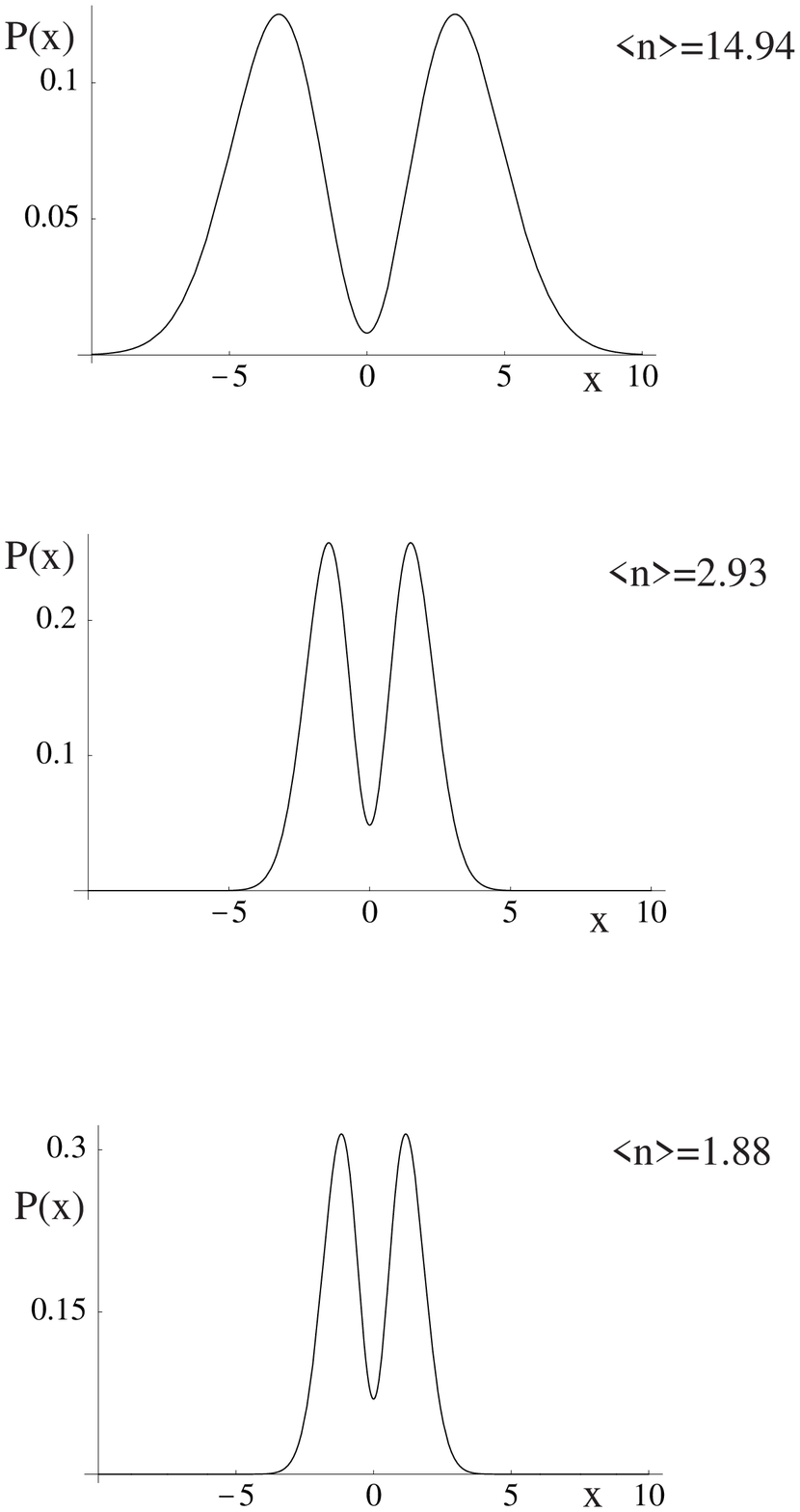,width=3.3in}}
\vspace{0.1cm}
\caption{\narrowtext
         Probability distributions $P(x)$ for the quadrature operator
         $x_{d_{+}}=(d_{+}^{\dagger}+d_{+})/\protect\sqrt{2}$ of the
         $d_{+}$ mode and for three values of $\langle n \rangle=N$.}
\label{fg:margx}
\end{figure}

\begin{figure}
\centerline{\psfig{figure=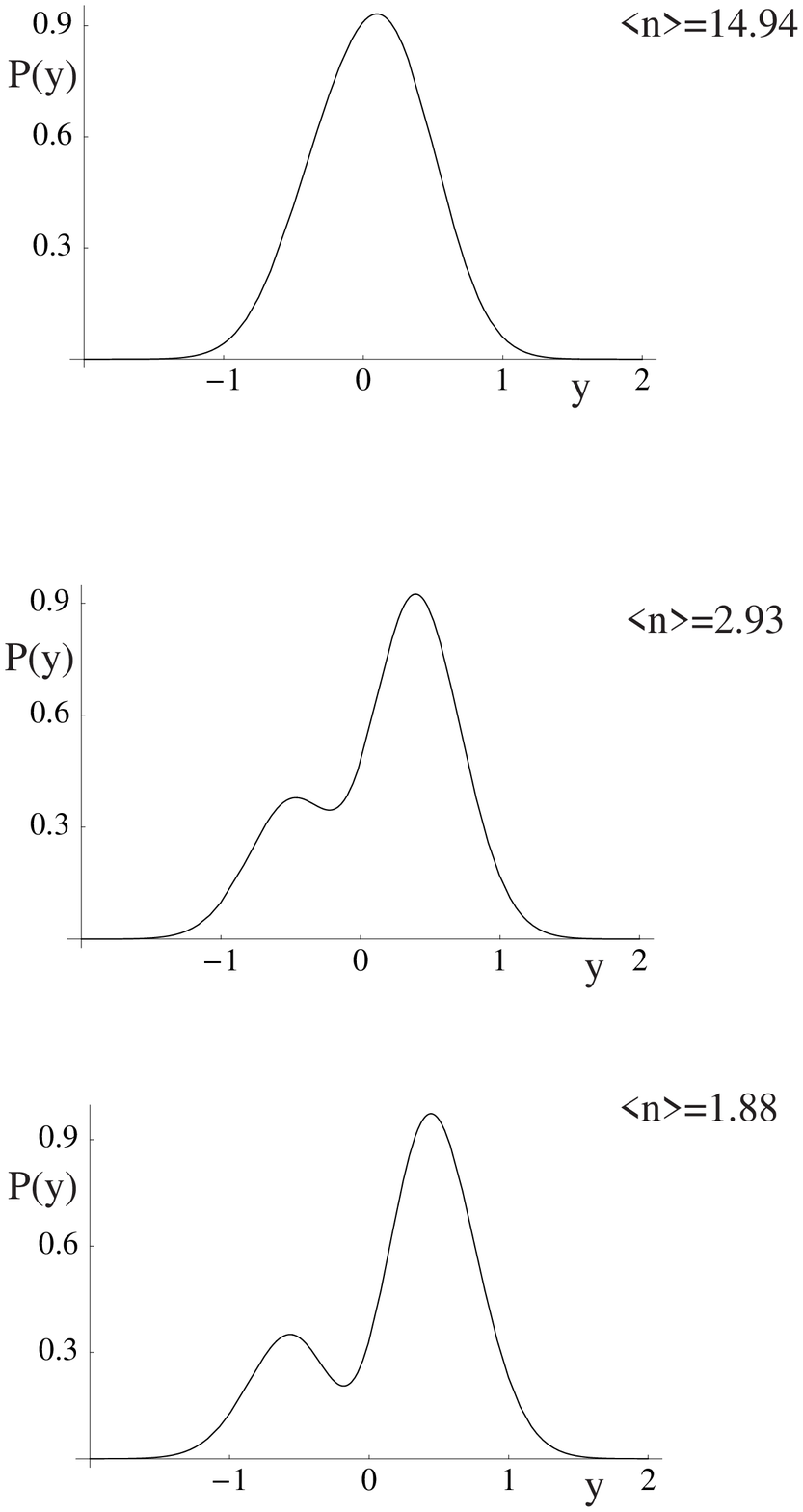,width=3.3in}}
\vspace{0.1cm}
\caption{\narrowtext
         Probability distributions $P(y)$ for the quadrature operator
         $y_{d_{+}}=i(d_{+}^{\dagger}-d_{+})/\protect\sqrt{2}$ of the
         $d_{+}$ mode and for three values of $\langle n \rangle=N$.}
\label{fg:margy}
\end{figure}

\begin{multicols}{2}
\vspace{-0.6cm}
\noindent
which is in very good agreement with the numerically computed exact 
one. As desired, the value of $W(x_{d_{+}},y_{d_{+}})$ at the origin 
may also be negative (depending on the parameters $\alpha$ and $\beta$
specifying the conditional measurement), reflecting the quantum properties
of the original 4-dimensional Wigner function~(\ref{eq:wigfour}).
Explicitly, one has
\begin{equation}
W(x_{d_{+}}=0,y_{d_{+}}=0) = 2 |\beta|^{2} 
\frac{2-\frac{\chi_{2}^{2}}{\kappa^{2}}}{4-\frac{\chi_{2}^{2}}{\kappa^{2}}}
- |\alpha|^{2}\;.
\label{eq:w00}
\end{equation} 
These results are graphically shown in 
Figs.~\ref{fg:wig1}--\ref{fg:margy}. In Figs.~\ref{fg:wig1} and 
\ref{fg:wig2} we have plotted the Wigner function~(\ref{eq:wd}) for
two values of $N$. In both figures two different viewpoints have been
selected for the tridimensional plots, in order to display most 
clearly the quantum superposition character of our cat-like state.
In particular, one should note that in both cases the Wigner function 
is negative around the origin. However, a comparison between
Fig.~\ref{fg:wig1} and Fig.~\ref{fg:wig2} shows that even though the two
Gaussian peaks are better separated for a larger number of photons, the
negativity of the Wigner function tends to disappear as soon as the initial
number of photons increases, as expected. This behavior is further 
confirmed by the inspection of the corresponding marginal distributions 
of the Wigner function~(\ref{eq:wd}), shown in Figs.~\ref{fg:margx} and
\ref{fg:margy}: $P(x)=P(x_{d_{+}})$ displays a larger separation 
between the peaks as the initial mean photon number $\langle n 
\rangle = N$ increases. On the other hand, $P(y)=P(y_{d_{+}})$ displays
the interference between the two macroscopic components, which tends
to be washed out when the number of photons increases. In fact, for 
$N=14.94$, the interference fringes have already disappeared.

\section{Discussion and Conclusions}
\label{conclu}

In this paper we have considered the generation of {\em entangled} 
Schr\"odinger-cat states in an optical parametric oscillator, as a 
relevant variant of the original proposal~\cite{kn:dem} which instead
had considered the amplifier case. In these works, the central point
(both conceptually and experimentally) is the
{\em quantum injection}~\cite{kn:dem} of the second nonlinear crystal 
with the output of the first parametric medium. In the present paper, 
we have computed the time evolution for the electromagnetic-field and 
chosen the initial condition needed for the  generation of the desired 
cat state. Such a state, however, lives in a eight-dimensional 
phase-space: therefore we have proposed three methods which are able 
to prove that it is an actual Schr\"odinger-cat state: direct
photodetection, measurement of the correlation functions, and 
measurement of the Wigner function. Our calculations show that the 
state produced in this way has indeed two macroscopic (mesoscopic) 
components---which are macroscopically (mesoscopically) 
distinguishable---and that they are in a coherent superposition (and 
not just in a statistical mixture), i.e. they display quantum 
interference.

A comparison with the performance of the corresponding OPA 
scheme~\cite{kn:dem} is in order here. First, the OPO has a larger 
conversion efficiency due to the enhancement factor of the parametric 
interaction, given by the presence of the cavities. This leads to a 
larger number of photon couples with the same pump power. Second, our 
Schr\"odinger-cat state is confined in the cavities, contrarily to 
what happens in the OPA case, where it is a traveling wave.
However, the price one has to pay in order to have these advantages, 
is given by the unavoidable cavity losses, that tend to destroy the 
coherence of the state when the number $N$ of initial photons tends to 
infinity. Such a phenomenon---decoherence~\cite{kn:zur,kn:prlha,kn:tra}---is
visualized by the 
progressive disappearance of the interference fringes and of the 
negativity of the Wigner function when $N$ increases. It is then clear 
that one has to consider a trade-off condition between the enhancement 
factor (a large $N$) and the losses (a low $\kappa$). This may lead to 
a comparison between the performances of the OPO and the 
OPA~\cite{kn:dem}: in particular, our OPO configuration is preferable 
when the mean number of initial photons $N$ [see 
Eq.~(\ref{eq:meannumb})] is larger than the corresponding 
parameter ($\sinh^{2}\chi t$~\cite{kn:dem}) of the OPA.

In conclusion, we think that an experiment along the lines outlined 
in this paper and in \cite{kn:dem}---which is realizable using 
presently available technology---is a promising candidate for 
producing entangled superpositions of macroscopically distinct quantum 
states.

\acknowledgments

It is a pleasure for us to acknowledge interesting and stimulating 
discussions with A.~Ekert and P.~Grangier.
This work has been partially supported by INFM (through the 1997 
Advanced Research Project ``Cat''), by the European Union
in the framework of the TMR Network ``Microlasers and Cavity QED'',
and by MURST through ``Cofinanziamento 1997''.

\appendix
\section*{}
\label{app}

The fact that the time $t$ during which we have the interaction within
the first nonlinear crystal is very short is of fundamental importance,
and it allows an immediate description of the experiment. To bring out
this most clearly, we develop an approximate treatment, which is 
however justified by the actual experimental values reported in
Ref.~\cite{kn:dem}.

The interaction time $t$, which is the time of flight of the photon 
within the first nonlinear crystal NL1, is given by
\begin{equation}
t=\frac{L_{k}n}{c}\simeq 10^{-11}\,{\rm sec}\,
\label{eq:intt}
\end{equation}
where $L_{k}$ is the crystal length, $n$ its refraction index, and
$c$ is the speed of light in vacuum. On the other hand, for an average 
pump power $P\simeq 300\, {\rm mW}$, the coupling strength is of the 
order of $\chi_{2}\simeq 6\cdot 10^{8}\, {\rm Hz}$.
In order to obtain ``macroscopic'' states, one needs a quite large 
initial mean number of photons in the parametric oscillator below 
threshold. This fixes the damping rates $\kappa_{2}=\kappa_{3}=\kappa$ to be 
slightly larger than $\chi_{2}$, since, from Eq.~(\ref{eq:meannumb}),
we have
\begin{equation}
\frac{\kappa^{2}}{\chi_{2}^{2}}=1+\frac{1}{2 \bar{N}}\;.
\label{eq:kchi}
\end{equation}
Therefore, we have $\kappa\simeq 6\cdot 10^{8}\, {\rm Hz}$, too. Since
the wavelength of the photon is $\lambda\simeq 7.3 \cdot 10^{-5}$ cm,
this amounts to having a {\em standard cavity}, with a quality factor
\begin{equation}
Q=\frac{2\pi c}{\lambda \kappa}\simeq 10^{5}\;.
\label{eq:qfactor}
\end{equation}
On the other hand, $\chi_{1}$ will be of the order of $\chi_{2}$.
In summary, we have
\begin{equation}
\chi_{2}t \simeq \chi_{1}t \simeq \kappa t\simeq 10^{-3}\;.
\label{eq:chit}
\end{equation}

From Eqs.~(\ref{eq:meq}--\ref{eq:hnl123}), (\ref{eq:infact}), and
(\ref{eq:rho123}), one has, for the time evolution of the combined
density matrix,
\begin{equation}
\rho_{123}(t)=e^{{\cal L}_{123}(t)}\rho_{23}(0)|0\rangle_{1}\langle 
0|\;.
\label{eq:ellero}
\end{equation}
Since ${\cal L}_{123} \propto \kappa $, $\chi_{1}$, $\chi_{2}$, it is
appropriate to expand the exponential $e^{{\cal L}_{123}t}$ in power
series up to second order in $\kappa t$, $\chi_{1}t$, $\chi_{2}t$, yielding
\begin{equation}
e^{{\cal L}_{123}t} \simeq 1 + {\cal L}_{123}t +
\frac{1}{2} {\cal L}_{123}^{2}t^{2}\;,
\label{eq:expand}
\end{equation}
and
\begin{equation}
{\cal L}_{123}\rho = {\cal L}_{23}\rho + 
\chi_{1}\left[a_{1}^{\dagger}a_{2}^{\dagger}-a_{1}a_{2}, \rho
\right]\;,
\label{eq:L123}
\end{equation}
where ${\cal L}_{23}$ is that part of the Liouvillian which only acts
on the modes $\vec{k}_{2}$ and $\vec{k}_{3}$, as given by
Eqs.~(\ref{eq:parmeq}) and (\ref{eq:hnl123}).
It is possible in this way to determine the conditional states
$\rho_{23}^{(0)}$, $\rho_{23}^{(1)}$, and the interference terms in
Eqs.~(\ref{eq:rhocon}--\ref{eq:rhocon23i}). We compute first
\begin{equation}
\rho_{2-3}^{(1)}\simeq\frac{a_{2}^{\dagger}\rho_{2-3}(0)a_{2}}{{\rm Tr}
(a_{2}^{\dagger}\rho_{2-3}(0)a_{2})}\;.
\label{eq:1rho1}
\end{equation}
The properties of this state are usually characterized by measuring 
the photon-number distribution of the mode 2 along direction 2. We 
have therefore to perform the trace over the mode 3 in 
Eq.~(\ref{eq:1rho1}), obtaining
\begin{equation}
\rho_{2}^{(1)\, {\rm red}}={\rm Tr}\left(\rho_{2-3}^{(1)}\right)
=\frac{a_{2}^{\dagger}\left({\rm Tr}_{3}\rho_{2-3}(0)\right)a_{2}}{{\rm Tr}
(a_{2}^{\dagger}\rho_{2-3}(0)a_{2})}\;.
\label{eq:rho21red}
\end{equation}
We already know that ${\rm Tr}_{3}\rho_{2-3}(0)$ is a thermal state 
with a mean number of photons given by $\bar{N}$ [see 
Eqs.~(\ref{eq:meannumb}), (\ref{eq:kchi}), and (\ref{eq:n20})], i.e.,
\begin{equation}
{\rm Tr}_{3}\rho_{2-3}(0)=\sum_{n=0}^{\infty} 
\left(\frac{\bar{N}}{1+\bar{N}}\right)^{n}|n\rangle\langle n|
\left(\frac{1}{1+\bar{N}}\right)\;,
\label{eq:thermal}
\end{equation}
[see Eq.~(\ref{eq:redtherm})] and consequently
\begin{eqnarray}
\rho_{2}^{(1)\, {\rm red}} & = & \sum_{n=0}^{\infty}P_{\rm H}(n) 
|n\rangle\langle n|\;,
\label{eq:phnn} \\
P_{\rm H}(n) & = & n\left(\frac{\bar{N}}{1+\bar{N}}\right)^{n-1}
\frac{1}{(1+\bar{N})^{2}}\;,
\label{eq:phn}
\end{eqnarray}
which is a sort of {\em shifted} thermal state and is identical to the
state obtained in the case  of the parametric amplifier~\cite{kn:dem} with
a mean number of photons given by Eq.~(\ref{eq:meannumb}).
On the other hand, we have, at the lowest order in $\chi_{1}t$,
\begin{mathletters}
\begin{eqnarray}
\rho_{2-3}^{(0)} & = & \langle 0|\rho_{1-2-3}(t) |0\rangle =
\left(1+{\cal L}_{23}t +\frac{{\cal L}_{23}^{2}t^{2}}{2}\right)
\rho_{2-3}(0)
\nonumber \\
 & & - \frac{\chi_{1}^{2}t^{2}}{2}
 \left(a_{2}a_{2}^{\dagger}\rho_{2-3}(0)
 +\rho_{2-3}(0)a_{2}a_{2}^{\dagger}\right)
\label{eq:0rho0a} \\
 & \simeq & \rho_{2-3}(0)\;,
\label{eq:0rho0b}
\end{eqnarray}
\end{mathletters}
and the state $\rho_{2}^{(0)\, {\rm red}}={\rm Tr}_{3}(\rho_{2-3}(0))$
conditioned upon the detection of no photons is essentially identical
to the initial usual thermal state.

\end{multicols}

\begin{multicols}{2}

\end{multicols}

\newpage

\begin{multicols}{2}

\end{multicols}

\begin{references}
\bibitem[*]{mau} Electronic address: mauro@camcat.unicam.it
\bibitem{kn:sc} E.~Schr\"odinger, Naturwiss. {\bf 23}, 807 (1935).
\bibitem{kn:cat} B.~Yurke and D.~Stoler, Phys. Rev. Lett. {\bf 57},
                 13 (1986). W.~Schleich, M.~Pernigo, and Fam Le Kien,
                 Phys. Rev. A {\bf 44}, 2172 (1991).  
\bibitem{kn:zur} W.~H.~Zurek, Phys. Rev. D {\bf 24}, 1516 (1981);
                 {\em ibid.} {\bf 26}, 1862 (1982);
                 Phys. Today {\bf 44}(10), 36 (1991), and
                 references therein.
\bibitem{kn:prlha} M.~Brune, E.~Hagley, J.~Dreyer, X.~Maitre, A.~Maali,
                   C.~Wunderlich, J.~M.~Raimond and S.~Haroche, Phys. Rev. Lett.
                   {\bf 77}, 4887 (1996).
\bibitem{kn:tra} S.~Habib, K.~Shizume, and W.~H.~Zurek,
                 Phys. Rev. Lett. {\bf 80}, 4361 (1998).
\bibitem{kn:win} C.~Monroe, D.~M.~Meekhof, B.~E.~King, and D.~J.~Wineland,
                 Science {\bf 272}, 1131 (1996).
\bibitem{kn:ent} E.~Schr\"odinger, Proc. Cambridge Philos. Soc. {\bf 31}, 
                 555 (1935).
\bibitem{kn:epr} A.~Einstein, B.~Podolsky, and N.~Rosen, Phys. Rev.
                 {\bf 47}, 777 (1935).
\bibitem{kn:tel} C.~Bennett, G.~Brassard, C.~Cr\'epeau, R.~Josza, A.~Peres,
                 and W.~K.~Wootters, Phys. Rev. Lett. {\bf 70},
                 1895 (1993).
                 D.~Bouwmeester, J.-V.~Pan, K.~Mattle, M.~Eibl, H.~Weinfurter,
                 and A.~Zeilinger, Nature (London) {\bf 390}, 575 (1997);
                 D.~Boschi, S.~Branca, F.~De~Martini, L.~Hardy, 
                 and S.~Popescu, Phys. Rev. Lett. {\bf 80}, 1121 (1998).
\bibitem{kn:cryp} A.~K.~Ekert, Phys. Rev. Lett. {\bf 67}, 661 (1991);
                  A.~K.~Ekert {\it et al.}, {\it ibid.} {\bf 69}, 
                  2881 (1992).
\bibitem{kn:qcomp} D.~P.~DiVincenzo, Science {\bf 270}, 255 (1995).
\bibitem{kn:dem} F.~De~Martini, Phys. Rev. Lett. {\bf 81}, 2842 (1998);
                 Phys. Lett. A {\bf 250}, 15 (1998).
\bibitem{kn:kwiat} P.~G.~Kwiat, K.~Mattle, H.~Weinfurter, A.~Zeilinger,
                   A.~V.~Sergienko, Y.~H.~Shih, Phys. Rev. Lett. {\bf 75},
                   4337 (1995). For the generation of an ultra-bright 
                   source of polarization-entangled photons also see:
                   P.~G.~Kwiat, E.~Waks, A.~G.~White, I.~Appelbaum, and
                   P.~H.~Eberhard, lanl e-print quant-ph/9810003.
\bibitem{kn:bell} J.~S.~Bell, Physics {\bf 1}, 195 (1964).
\bibitem{kn:ghz} D.~M.~Greenberger, M.~A.~Horne, and A.~Zeilinger,
                 Am. J. Phys. {\bf 58}, 1131 (1990);
                 N.~D.~Mermin, Phys. Rev. Lett. {\bf 65}, 1838 (1990). 
                 D.~Bouwmeester {\it et al.}, to appear (1998).
\bibitem{kn:shih} M.~H.~Rubin, D.~N.~Klyshko, Y.~H.~Shih, A.~V.~Sergienko,
                  Phys. Rev. A {\bf 50}, 5122 (1994).
\bibitem{kn:haro} L.~Davidovich, M.~Brune, J.~M.~Raimond and S.~Haroche,
                  Phys. Rev. A {\bf 53}, 1295 (1996).
\bibitem{kn:mandel} X.~Y.~Zou, L.~J.~Wang, L.~Mandel, Phys. Rev. Lett. {\bf 67},
                    318 (1991); L.~J.~Wang, X.~Y.~Zou, L.~Mandel, Phys. Rev. A
                    {\bf 44}, 4614 (1991); T.~P.~Grayson, X.~Y.~Zou, D.~Branning,
                    J.~R.~Torgerson, L.~Mandel, Phys. Rev. A {\bf 48},
                    4793 (1993). 
\bibitem{kn:leg2} A.~O.~Caldeira and A.~J.~Leggett, Phys. Rev. A {\bf 31},
                  1059 (1985); D.~F.~Walls and G.~J.~Milburn, Phys. Rev. A
                  {\bf 31}, 2403 (1985).
\bibitem{kn:parosc} M.O.~Scully and M.S.~Zubairy, {\em Quantum Optics}
                    (Cambridge University Press, Cambridge, 1997).
                    G.~J.~Milburn and D.~F.~Walls, {\em Quantum Optics} 
                    (Springer-Verlag, Berlin, 1994).
\bibitem{kn:cond} B.~Sherman and G.~Kurizki, Phys. Rev. A {\bf 45}, R7674
                  (1992); B.~Sherman, H.~Moya-Cessa, P.~L.~Knight,
                  and G.~Kurizki, {\it ibid.} {\bf 49}, 535 (1994).
                  K.~Vogel, V.~M.~Akulin, and W.~P.~Schleich,
                  Phys. Rev. Lett. {\bf 71}, 1816 (1993).
\bibitem{kn:wig} M.~Hillery, R.~F.~O'Connell, M.~O.~Scully, and E.~P.~Wigner,
                 Phys. Rep. {\bf 106}, 121 (1986). 
\bibitem{kn:gar} C.~W.~Gardiner, {\em Quantum Noise}
                 (Springer-Verlag, Berlin,1991).
\bibitem{kn:risk} H.~Risken, {\em The Fokker--Planck Equation}
                  (Springer-Verlag, Berlin, 1989).
\bibitem{kn:miwa} G.~J.~Milburn and D.~F.~Walls, {\em Quantum Optics}
                  (Springer-Verlag, Berlin, 1991).
\end{references}
\end{document}